\documentclass{aa}
\usepackage{psfig}

\newcommand{\MC}{\multicolumn}
\newcounter{qub}
\newcommand{\qq}{\addtocounter{qub}{1}\arabic{qub}}

\begin{document}
\thesaurus{20(04.19.1; 11.06.2; 11.04.1; 11.19.3; 11.03.2)}
\title{ The Hamburg/SAO Survey for Emission--Line Galaxies }
\subtitle{ III. The Third List of 81 Galaxies }

\author{U.~Hopp\inst{1}\fnmsep\inst{9}
\and D.~Engels\inst{2}
\and R.F.~Green\inst{3}
\and A.V.~Ugryumov\inst{4}
\and Y.I.~Izotov\inst{5}\fnmsep\inst{10}
\and H.-J.~Hagen\inst{2}
\and A.Y.~Kniazev\inst{4}
\and V.A.~Lipovetsky\inst{4}\fnmsep\thanks{Deceased 1996 September 22.}
\and S.A.~Pustilnik\inst{4}
\and N.~Brosch\inst{6}
\and J.~Masegosa\inst{7}
\and J.-M.~Martin\inst{8}
\and I.~M\'arquez\inst{7}
}

\offprints{hopp@usm.uni-muenchen.de}

\institute{
Universit\"atssternwarte M\"unchen, Scheiner Str. 1, D-81679 M\"unchen, Germany
\and Hamburger Sternwarte, Gojenbergsweg 112, D-21029 Hamburg, Germany
\and National Optical Astronomy Observatories, Tucson, AZ, 85726-6732, USA
\and Special Astrophysical Observatory, Nizhnij Arkhyz, Karachai-Circessia,
357147, Russia
\and Main Astronomical Observatory, Goloseevo, Kiev-127, 03680, Ukraine
\and Wise Observatory, Tel-Aviv University, Tel-Aviv 69978, Israel
\and Instituto de Astrofisica de Andalucia, CSIC, Aptdo. 3004, 18080, Granada, Spain
\and D\'epartement de Radioastronomie ARPEGES, Observatoire de Paris, F-92195 Meudon Cedex, France
\and Visiting astronomer at Calar Alto Observatory, Spain
\and Visiting astronomer at Kitt Peak Observatory, USA
}

\date{Received \hskip 2cm; Accepted}

\maketitle

\markboth{U.Hopp et al.: The Hamburg/SAO Survey for Emission-Line
Galaxies. III}
{The Hamburg/SAO Survey. III}

\begin{abstract}

We present the third list with results\footnote{ Tables 2 to 6 are
available only in electronic form at the CDS via anonymous ftp to
cdsarc.u-strasbg.fr (130.79.128.5) or via
http://cdsweb.u-strasbg.fr/Abstract.html. Figures A1 to A9 will be
made available only in the electronic version of the journal.}  of the
Hamburg/SAO Survey for Emission-Line Galaxies (HSS therein, SAO --
Special Astrophysical Observatory, Russia). This survey is based on
the digitized objective-prism photoplate database of the Hamburg
Quasar Survey (HQS). 

Here, we present new spectroscopic results of candidates which were
obtained in 1998 with the 2.1\,m KPNO and the 2.2\,m Calar Alto
telescopes.  All candidates are selected in the declination band
+35$^{\circ}$ to +40$^{\circ}$.

The follow-up spectroscopy with the 2\,m class telescopes confirmed 
85 emission-line objects out of 113 observed candidates and allowed their
quantitative spectral classification. 
For 80 of them, the redshifts are determined for the first time.
For 5 previously known ELGs, line ratios are presented for the first time.
We could classify 55 out of the 85 emission-line objects as BCG/H{\sc
ii} galaxies or probable BCGs, 4 -- as QSOs, 6 -- as Seyfert galaxies, 
1 -- as super-association in a subluminous spiral galaxy,
and 11 are low-excitation objects -- either starburst nuclear (SBN),
or dwarf amorphous nuclei starburst galaxies (DANS). We could 
not classify 8 ELGs. Further, for 8 more galaxies we did not detect any 
significant emission lines. 


\end{abstract}


\begin{table*}
\begin{center}
\caption{\label{Tab1} Journal of observations}
\begin{tabular}{crllccc} \\ \hline
\MC{1}{c}{ Date } &
\MC{1}{c}{ Telescope } &
\MC{1}{c}{ Instrument } &
\MC{1}{c}{ Grating } &
\MC{1}{c}{ Wavelength } &
\MC{1}{c}{ Dispersion } &
\MC{1}{c}{ Observed } \\

\MC{1}{c}{ } & & &
\MC{1}{c}{ [\AA/mm] } &
\MC{1}{c}{ range [\AA] } &
\MC{1}{c}{ [\AA/pixel] } &
\MC{1}{c}{ number } \\

\MC{1}{c}{ (1) } &
\MC{1}{c}{ (2) } &
\MC{1}{c}{ (3) } &
\MC{1}{c}{ (4) } &
\MC{1}{c}{ (5) } &
\MC{1}{c}{ (6) } &
\MC{1}{c}{ (7) } \\
\hline
\\[-0.3cm]
22.05-25.05.1998 & 2.1~m KPNO &  GoldCam & 165  & 3600--7400 & 2.7 & 69 \\
27.05-01.06.1998 & 2.2~m CAHA & CAFOS & 187 & 3700--8100 & 8.9 & 44 \\

\hline \\[--0.2cm]

\end{tabular}
\end{center}
\end{table*}

\section{Introduction}

Objective prism surveys for emission-line galaxies (ELGs) are the main
source of large samples of both AGNs and galaxies with enhanced star
formation (SF) activity. Several large samples of ELGs were published
since the end of the 1980s. They include the samples of the University
of Michigan (UM) survey (Salzer \& MacAlpine \cite{Salzer88}; Salzer
\cite{Salzer89}; Salzer et al.  \cite{Salzeretal89}) near the equator,
the Tololo and C\'alan-Tololo survey samples (Terlevich et
al. \cite{Terlevich91}; Maza et al.  \cite{Maza91}) and the recent
Marseille Schmidt survey (Surace \& Comte \cite{Surace98}) of the
Southern sky.

In the Northern sky, large samples of ELGs have appeared during the
last decade thanks to such objective prism surveys as the First and
the Second Byurakan (SBS) surveys (Markarian et
al.~\cite{Markarian83}; Izotov et al. \cite{Izotov93a};
Stepanian~\cite{Stepanian94}; Pustilnik et al.  \cite{Pustilnik95}),
the Case survey (Pesch et al. \cite{Pesch95}; Salzer et al.
\cite{Salzer95}; Ugryumov et al. \cite{Ugryumov98}), and the
Heidelberg void survey (Popescu et al. \cite{Popescu96},
\cite{Popescu97}, \cite{Popescu98}). All these projects employed
detection of strong emission lines on blue-sensitive photoplates. A
complementary approach was based on the search of strong
H$\alpha$-emission on red objective prism plates as e.g. in the
Universidad Complutense de Madrid (UCM) survey (Zamorano et
al. \cite{Zamorano94}; Zamorano et al. \cite{Zamorano96}; Gallego et
al. \cite{Gallego97}), and the MBC (Montreal) survey (Coziol et al.
\cite{Coziol93}, \cite{Coziol94}).

Despite the large effort to establish the above mentioned surveys,
they yielded only relatively small {\it complete} samples on the order
of 10$^2$ blue compact galaxies (BCGs hereafter, e.g. Thuan et al.,
1999). This is related to the relatively low surface density of the
objects in the surveys of about 0.2-0.3 per sq. deg. (e.g. Popescu et
al. \cite{Popescu97}). But only complete samples of sufficient size
will allow studying the distribution of the inherent physical
parameters of BCGs.  The experience of all these surveys can be
summarized as follows. To push progress in statistical studies of
low-mass galaxies with star formation bursts, a reasonably large
volume has to be surveyed and the selection has to be done by well
understood selection procedures. Especially for BCGs of extremely low
chemical abundances, which seem to be very rare objects in the local
universe after all we have learned so far, a coverage of several
10$^3$ square degrees down to the technical limits of the surveys is
essential. These limits are at magnitudes as faint as $m_{\rm b}$ = 18
-- 19 mag.  To derive a statistically robust sample of sufficient
sizes from a very large field survey, objective selection procedures
for the ELGs have to be applied.

With the data described below and in papers I and II (Ugryumov et
al. \cite{Ugryumov99}; Pustilnik et al. \cite{Pustilnik99}) of this
series, the authors pursue the goal of creating a new large sample of
H{\sc ii} galaxies, or BCGs in a zone with a total area of the order
1500 square degrees.  This region will fill the gap between the zones
of the SBS and the region covered by the Case survey. The SBS is
situated at $\alpha = 7^{\rm h} 40^{\rm m} \div 17^{\rm h} 20^{\rm
m}$, $\delta = +49\degr \div +61\degr$, while the Case covers $\alpha
= 8^{\rm h} 00^{\rm m} \div 16^{\rm h} 20^{\rm m}$, $\delta = +29\degr
\div +38\degr$. For a description of the BCGs found in these two
surveys, see Izotov et al. (\cite{Izotov93a}, \cite{Izotov93b}), Thuan
et al. (\cite{Thuan94}) and Pustilnik et al.  (\cite{Pustilnik95}) for
the SBS and Salzer et al. (\cite{Salzer95}), Ugryumov
(\cite{Ugryumov97}), Ugryumov et al. (\cite{Ugryumov98}) for the Case
survey which is still in progress.

Thus, the new Hamburg/SAO Survey (HSS) for emission-line galaxies
leads, firstly, to the creation of a new BCG/H{\sc ii} galaxy sample
in a large sky region with the boundaries $7^{\rm h} 20^{\rm m}$ to
$17^{\rm h} 40^{\rm m}$ in right ascension and $+35\degr$ to
$+50\degr$ in declination.  Secondly, after combining the three BCG
samples in the SBS, the Case and the HSS zones, a large Northern BCG
sample covering about 3000 square degrees will be available. The main
goal of the project is the search for emission-line galaxies (ELG) in
order to create a new deep sample of blue compact/H{\sc ii} galaxies
(BCG) in a large area of the sky. Another important goal of this work
is to search for new extremely low-metallicity galaxies.

This is the third article of a series devoted to follow-up
spectroscopy results of HSS ELG candidates. It deals with 113
candidates selected in the band between $+35\degr$ and $+40\degr$ in
declination which is complementary to the zone
+40$^{\circ}$$\div$+50$^{\circ}$ studied in our previous papers.  The
basic ideas of the HSS and its selection methods of ELG candidates are
described along with the first results of the follow-up spectroscopy
in Ugryumov et al. (\cite{Ugryumov99}) (Paper~I). The final selection
was slightly modified to improve significantly the detection rate of
ELGs in follow-up spectroscopy as described in Paper~II. In short, the
ELG candidate selection criteria applied are a blue or flat continuum
(near $\lambda$~4000~\AA) and the presence of strong or moderate
[O{\sc iii}]\,$\lambda\lambda$\,4959,5007~\AA\ emission lines
recognized on digitized prism spectra. Candidates accepted are
restricted to the $B$-magnitude range $16^{\rm m} - 19\fm5$.

The article is organized as follows. In section 2 we give the details
of the spectroscopic observations and of the data reduction. In
section 3 the results of the observations are presented in several
tables. Along with general parameters for the emission-line galaxies
and several quasars, the parameters of the strongest emission lines of the
ELGs are summarized in a separate table. The information on two
non-emission-line galaxies is presented as well.  In section 4 we
briefly discuss the new data and summarize the current state of the
Hamburg/SAO survey. Throughout this paper a Hubble constant $H_0$ = 75
km$\,$s$^{-1}$ Mpc$^{-1}$ is used.

\section{Spectral observations and data reduction}

All results presented below were obtained by observations in a snap-shot
mode during two runs with the KPNO 2.1\,m telescope and
the Calar Alto 2.2\,m telescope (see Table~1).

\subsection{Observations with the KPNO 2.1m telescope}

The observations were made with the GoldCam spectrograph used in
conjunction with the 3K$\times$1K CCD detector. We used a
2$''$$\times$229$''$ slit with the grating 09 (316 grooves mm$^{-1}$)
in its first order, and a GG 375 order separation filter.  This filter
cuts off all second-order contamination for wavelengths blueward of
7400\AA\, which is the wavelength region of interest here. This
instrumental set-up gave a spatial scale along the slit of 0\farcs75
pixel$^{-1}$, a scale perpendicular to the slit of 2.7\AA\
pixel$^{-1}$, a spectral range of 3700--7500\AA\, and a spectral
resolution of $\sim$ 5\AA. These parameters permitted cover
simultaneous coverage of the blue and red spectral range with all the lines of
interest in a single exposure and with enough spectral resolution to
separate important emission lines such as H$\gamma$\ $\lambda$4340 and
[O{\sc iii}] $\lambda$4363, and H$\alpha$\ $\lambda$6563 and [N{\sc
ii}] $\lambda$6584. Normally, short exposures were used (5 minutes) in
order to detect strong emission lines, to measure redshifts and make
some crude classification.

Reference spectra of an Ar--Ne--He lamp were recorded to provide a
wavelength calibration. The spectrophotometric standard star Feige 34
from Massey et al. (\cite{Massey88}) was observed for the flux
calibration at least once a night. No effort was made to orient the
slit along the parallactic angle, so line flux ratios could be
spectrophotometrically inaccurate. The observations were complemented by
dome flats, bias-, and dark frames.  The seeing was about 3$\arcsec$
(FWHM).

\subsection{Calar Alto 2.2m telescope observations}

Follow-up spectroscopy with this telescope was conducted as a back-up
for a main program which needed photometric conditions. So, the
observations presented here were obtained in non-photometric
conditions and the absolute flux calibration of the data is
unreliable.

The Cassegrain focal reducer CAFOS of the 2.2m telescope was used with
a long slit of 300$\arcsec \times$ 3$\arcsec$ and a grism of 187 \AA\,
mm$^{-1}$ linear dispersion. Spectra were recorded on a 2K$\times$2K
Site CCD operated in a 2$\times$1 binned mode (binning only along the
dispersion direction), resulting in a spectral resolution of about 20
\AA\, (FWHM), and a wavelength coverage {\bf $\lambda = 3700 - 8100$
}~\AA.  No order separation filter was installed. The slit orientation
was again not aligned with the parallactic angle to keep the
duty-cycle high.  The exposure times varied between 10 and 15 minutes
depending on the object brightness. The observations were complemented
by standard star flux measurements, Hg--He--Cd lamp exposures for
wavelength calibration, dome flat-, bias-, and dark-frames. The seeing
was between 1.5 and 2.5 $\arcsec$ (FWHM).

\subsection{Data reduction}

\subsubsection{Reduction of the KPNO 2.1m telescope data}

The KPNO two-dimensional spectra were bias subtracted and flat-field
corrected.  We then use the IRAF\footnote[2]{IRAF is distributed by
National Optical Astronomical Observatories, which is operated by the
Association of Universities for Research in Astronomy, Inc., under
cooperative agreement with the National Science Foundation} software
routines IDENTIFY, REIDENTIFY, FITCOORD, TRANSFORM to do the
wavelength calibration and the correction for distortion and tilt for
each frame. Then the one-dimensional spectra were extracted from each
frame using the APALL routine without weighting. For all objects we
extracted the brightest part of the galaxy covering a spatial size of
7\arcsec.  All extracted spectra from the same object were then
co-added. Cosmic ray hits have been removed manually.  To derive the
instrumental response function, we have fitted the observed spectral
energy distribution of the standard star Feige 34 with a high-order
polynomial.
%

\subsubsection{Reduction of the Calar Alto 2.2m telescope data}

This reduction was fully done at SAO with the standard reduction
system MIDAS (Munich Image Data Analysis System, Grosb{\o}l
\cite{Grosbol89}). We applied the context LONG as follows: bias and
dark subtraction, flat-fielding, cosmic-ray removal.  After the
wavelength mapping, a night sky 2-D background subtraction was
performed.  1-D spectra were extracted by adding the consecutive CCD
rows centered on the object intensity peak along the slit.  Then the
corrections for atmospheric extinction and flux calibration were
applied. Despite of the non-photometric observing conditions, we
corrected the spectra for the instrumental response with a response
curve established by observations of the spectrophotometric standard
star \mbox{BD+33$^{\circ}$ 2642}.

\subsubsection{Line parameter measurements}

In the final spectra, redshifts and line fluxes are measured within
MIDAS, applying Gaussian fitting to the emission lines. To determine
redshifts for individual galaxies, averages are taken over the
prominent individual emission lines (mostly H$\beta$, H$\alpha$,
[O{\sc iii}]\,$\lambda$\,4959,5007~\AA).  The line [O{\sc
ii}]\,$\lambda$\,3727~\AA\ is not included in the redshift determination
since for most of the objects its observed wavelength is determined
with significantly larger uncertainties due to the extrapolation of
the linear scale below the first line of the reference spectrum 
(He{\sc i}\,$\lambda$\,3889~\AA\ ). However [O{\sc ii}]\,$\lambda$\,3727~\AA\
was used to determine the redshift in rare cases where it is the only
strong emission line. The errors of the redshift in such cases can be
several times larger than the typical one (compare Table 2).

To improve the accuracy of the redshift determination for the Calar
Alto spectra, and further, to reduce possible small systematic shifts in the
zero point of the wavelength calibration, we additionally checked the
wavelengths of night sky emission lines on the 2-D spectra at the
position of the object spectrum.  If some measurable shift was
detected it was incorporated in measurements of emission line
positions. 

The emission line fluxes are computed by summing up the pixel
intensities inside the line region applying standard MIDAS program
tools.  For all spectra, the individual emission line fluxes of the
H$\alpha$, [N{\sc ii}]\,$\lambda\lambda$\,6548,6583~\AA\ and [S{\sc
ii}]\,$\lambda\lambda$\,6716,6731~\AA\ line blends are obtained by
summing up pixel intensities over the total blend and then modeling
the individual line fluxes using Gaussian fitting.

\section{Results of follow--up spectroscopy}

In total 108 new candidates and 5 known ELGs have been observed.
Among them, 81 are new or confirmed emission-line galaxies, 4 are quasars
(all with redshifts in the range 3.07 to 3.20), and 8 are galaxies
without emission lines. Only 2 of the latter have good enough 
S/N ratio to identify absorption features enabling measurements of their
redshifts. The remaining 20 objects appeared to be either stars with
characteristic absorption lines or stellar objects with featureless
spectra where the signal-to-noise ratio was insufficient to identify lines.

\subsection{Emission-line galaxies}

The new emission line galaxies are listed in Table 2 containing 
the following information: \\
 {\it column 1:} The object's IAU-type name with the prefix HS. We note
by asterisk objects observed at Calar Alto. \\
 {\it column 2:} Right ascension for equinox B1950. \\
 {\it column 3:} Declination for equinox B1950.
The coordinates were measured on direct plates of the HQS
and are accurate to $\sim$ 2$\arcsec$ (Hagen et al. \cite{Hagen95}). \\
 {\it column 4:} Heliocentric velocity and its r.m.s. uncertainty in
km~s$^{-1}$. \\
 {\it column 5:} Apparent $B$-magnitude obtained by calibration of the digitized
photoplates with photometric standard stars (Engels et al. \cite{Engels94}),
having an r.m.s. accuracy of $\sim$ $0\fm5$ for objects fainter than
$m_{\rm B}$ = $16\fm0$ (Popescu et al. \cite{Popescu96}).
Since the algorithm to calibrate the objective prism spectra is
optimized for point sources the brightnesses of extended galaxies are
underestimated. The resulting systematic uncertainties are expected to
be as large as 2 mag (Popescu et al. \cite{Popescu96}). For about 1/3
of our objects, $B$-magnitudes are unavailable at the moment. We present
for them blue magnitudes obtained from the APM database. They are
marked by a ``plus" before the value in the corresponding
column. According to our estimate they are systematically brighter by
$0\fm92$ than the $B$-magnitudes obtained by calibration of the
digitized photoplates (r.m.s.  $1\fm02$). \\
 {\it column 6:} Absolute $B$-magnitude, calculated from the apparent
$B$-magnitude and the heliocentric velocity. No correction for galactic
extinction is made as all objects are located at high
galactic latitudes and because the corrections are significantly smaller
than the uncertainties of the magnitudes. \\
 {\it column 7:} Preliminary spectral classification type according to
the spectral data presented in this article. BCG means that the galaxy
posesses a characteristic H{\sc ii}-region spectrum and that the
luminosity is low enough. SBN and DANS are galaxies of lower
excitation with a corresponding position in line ratio diagrams, as
discussed in Paper~I. SBN are the brighter fraction of this type. We
here follow the notation of Salzer et al. (1989). Seyfert galaxies are
separated mainly on diagnostic diagrams as AGN. But if their emission
lines are quite narrow, they probably should be classified as Sy2. SA
is a probable super-association at the rim of an edge-on nearby disc
galaxy.  Six objects are difficult to classify. They are coded as
NON. \\
 {\it column 8:} One or more alternative names, according to the
information from NED.\footnote{
NED is operated by the Jet Propulsion Laboratory, California
Institute of Technology, under contract with the National Aeronautics
and Space Administration.}

The spectra of all emission-line galaxies are shown in Appendix~A,
which is available only in the electronic version of the journal.

The results of line flux measurements are given in Table~4.
It contains the following information: \\
 {\it column 1:} The object's IAU-type name with the prefix HS.
By asterisk we note the objects observed  during
non-photometric conditions. \\
 {\it column 2:} Observed flux (in
10$^{-16}$\,erg\,s$^{-1}$\,cm$^{-2}$) of the H$\beta$\, line. For the
few objects without an
H$\beta$ emission line the fluxes are given for H$\alpha$ marked by a
``plus''.  For the objects observed on Calar Alto during
non-photometric conditions this parameter is unreliable and marked by
(:). \\
 {\it columns 3,4,5:} The observed flux ratios [O{\sc ii}]/H$\beta$,
[O{\sc iii}]/H$\beta$ and H$\alpha$/H$\beta$. \\
 {\it columns 6,7:} The observed flux ratios
[N{\sc ii}]\,$\lambda$\,6583~\AA/H$\alpha$, and
([S{\sc ii}]\,$\lambda$\,6716~\AA\ + \,$\lambda$\,6731~\AA)/H$\alpha$. \\
 {\it columns 8,9,10:} Equivalent widths of the lines
[O{\sc ii}]\,$\lambda$\,3727~\AA, H$\beta$ and
[O{\sc iii}]\,$\lambda$\,5007~\AA.
For the few objects without a detected H$\beta$ emission line the equivalent
widths are given for H$\alpha$ marked by a ``plus''.
\\

\noindent
Below we give notes on several individual objects: 

\noindent
{\it HS1015+3717}: In the spectrum of this object a cosmic ray hit is
exactly on the line [O{\sc iii}]$\lambda$4959\AA. This was not
corrected in the figure shown in Appendix~A. \\
{\it HS1214+3801}: This is seemingly a supergiant H{\sc ii}-region at
the very rim of the nearby edge-on disc galaxy (SA(s)cd) NGC~4244
($V_{\rm hel}$ = 224~km\,s$^{-1}$ and $B_{\rm T}$=10.88). At the accepted
distance of NGC~4244 ($D$ = 4.5 Mpc) $M_{\rm B}$ of HS1214+3801 is about
--$11\fm8$. The difference between the systemic radial velocity of the
host galaxy and H{\sc ii}-region is small (32~km\,s$^{-1}$) and does
not contradict that HS1214+3801 belongs to NGC~4244.  However,
the velocity field of NGC~4244 near the
position of the H{\sc ii}-region is unkown. Both, the single-dish H{\sc
i}-measurements as summarized in Huchtmeier \& Richter
(\cite{Huchtmeier89}), and an estimate of the maximum rotational velocity
$V_{\rm rot} \leq$ 130~km\,s$^{-1}$ (which we obtained through the
Tully-Fisher relation from the absolute $B$-band magnitude of NGC~4244
of $\sim$ --$17\fm8$), yield a range of expected velocity differences
between the galaxian material and HS~1214+3801 of up to +160 or
--100~km\,s$^{-1}$.  But since the 2-D spectrum of HS~1214+3801 with a
total spatial extent of about 20$^{\prime\prime}$ ($\approx$0.5
kpc) shows evidence of internal motions with an amplitude of about
50~km\,s$^{-1}$ we need to consider an alternative interpretation for
this object as a companion BCG. Its SF burst may be triggered due to
the tidal effect from the more massive galaxy, similar to the case of
HS~1717+4955 described in Kniazev et al. (\cite{Kniazev2000}).  To
check this option one needs a detailed map of the NGC~4244 velocity field
including HS~1214+3801.\\
{\it HS1214+3922}: This BCG was reobserved with higher S/N ratio
in order to measure the flux of the [O{\sc iii}]\,$\lambda$\,4363 line,
necessary to determine unambiguously the electron temperature
$T_e$([O{\sc iii}]) of the H{\sc ii}-region and the oxygen abundance.
A preliminary determination according to the procedure described by
Izotov et al. (\cite{Izotov97}) shows that it has
the low oxygen abundance of  log(O/H)~+~12~=~7.76.

\subsection{Quasars}

In the course of our follow-up spectroscopy, four QSOs were discovered
with a strong emission line in the wavelength region between 5000~\AA\
and the sensitivity break of the Kodak IIIa-J photoemulsion near
5400~\AA.  In all of them, we identified Ly$\alpha$\,$\lambda$\,1216
redshifted to $z$ $\sim$ 3 as the responsible line. This strong line
produces an easily visible emission peak in the digitized prism
spectra even for very faint objects ($B$ $\sim 19\fm0 - 20\fm0$) which
is hard to distinguish from low-redshift [O{\sc iii}] features. 
Else, QSOs were not selected as candidates for follow-up spectroscopy.

The data for these four new high-redshift quasars are presented in
Table~3.  Finding charts and plots of their spectra can be found
on the www-site of the Hamburg Quasar Survey
(http://www.hs.uni-hamburg.de/hqs.html).

\subsection{Non-emission-line objects}

In total, for 28 candidates no (trustworthy) emission lines are detected.
We divided them into three categories.

\subsubsection{Absorption-line galaxies}

For two bright non-ELG galaxies the signal-to-noise ratio
of spectra was sufficient to detect absorption lines, allowing
the determination of redshifts. The data are presented in Table~5.

\subsubsection{Stellar objects}

To separate the stars among the objects missing detectable emission
lines we cross-correlated a list of the most common stellar features
with the observed spectra.  In total, 13 objects with definite stellar
spectra and redshifts close to zero were identified. Four of them are
obvious K or M-stars.  The rest were classified roughly in categories
from definite A-stars to F or G-stars, with most of them intermediate
between F and G. The data for these stars are presented in Table~6.

\subsubsection{Non-classified  objects}

Thirteen non emission-line objects are hard to classify at all. Their
continua have too low signal-to-noise ratio to detect trustworthy
absorption features, or the equivalent width of the emission lines 
is too small. Six
of them are certainly non-stellar on DSS images, and classified as
well as non-stellar in the APM database. From our spectra in the range
$\approx$~4000 to $\approx$~7300--8000~\AA, we can exclude the
presence of strong H$\alpha$.  The remaining 7 objects are
indistinguishable from stellar ones, and we suggest that most of them
are galactic stars.  One of the galaxies -- namely HS 1232+3609, was
presented after our observations in the paper by Popescu et al.
(\cite{Popescu98}) as an ELG with $z$=0.2529. Our spectrum is too noisy,
and we could not identify any significant emission with this redshift.

\section{Discussion}

Altogether we have observed 113 objects preselected as ELG candidates
on HQS objective prism plates, of which 108 had no previous
spectroscopic information. Of those 85 objects (75~\%) are found to
be either ELGs, or quasars. Of 81 detected ELGs, 55 were classified
based on the character of their spectra and their absolute magnitudes
as H{\sc ii}/BCGs or probable BCGs.  According to their line intensity
ratios, six galaxies are of the Sy type, five of them probably of type
Sy2, while the continuum bump blueward to H$\beta$ in the spectrum of
HS~1526+3729 suggests an identification as the Fe II emission line blend
typical for Sy1 galaxies. But we caution the relatively low S/N ratio
in that part of the spectrum.  One very faint object (HS~1214+3801) of
absolute magnitude $M_{\rm B}= -11\fm8$ is probably a
super-association in the dwarf spiral NGC~4244, or a BCG companion to
this subluminous spiral galaxy.  Eight candidates are difficult to
classify. The remaining 11 ELGs are objects of lower excitation:
either starburst nuclei galaxies (SBN and probable SBN) or their lower
mass analogs, dwarf amorphous nuclear starburst galaxies (DANS or
probable DANS).  Since the main goal of the HSS is an efficient search
for new BCGs, the fraction of this type among all new detected ELGs
($\sim$68~\%, or 65~\% among all emission-line objects) is
encouraging.

The distributions of the new HSS ELGs in the line-ratio diagrams
[O{\sc iii}]\,$\lambda$\,5007/H$\beta$ versus
[N{\sc ii}]\,$\lambda$\,6583/H$\alpha$ and
[O{\sc iii}]\,$\lambda$\,5007/H$\beta$ versus
[O{\sc ii}]\,$\lambda$\,3727/[O{\sc iii}]\,$\lambda$\,5007
(see Baldwin et al. (\cite{Baldwin81}), Veilleux \& Osterbrock
(\cite{Veilleux87}) for details) in general are similar
to those shown in Paper~I.

Compared to Paper~II, we picked up significantly fewer
low-luminosity ELGs ($M_{\rm B} \leq -15$). This is partly
connected to the modest size of the telescopes used (2m versus 6m in
Paper~II) and the range of apparent brightness of ELG
candidates observed for this paper. Probably more important is that
the Calar Alto observations prefered apparently bright objects (only 4
fainter than 18.5) due to the back-up status of the measurements which
was prompted by a modest weather quality.

Altogether in Papers I through III, we discovered 257 new
emission-line objects (14 of them QSOs), and for 35 more galaxies we got
quantitative data for their emission lines.
Preliminary classification of the 278 ELGs yields 206 confident or probable
blue compact/low-mass H{\sc ii} galaxies. Thus a large fraction of BCGs
relative to all ELGs is found ($\sim$~74~\%) demonstrating the high
efficiency of this survey to find galaxies with H{\sc ii}-type spectra
on the Hamburg Quasar Survey photoplates. A statistical analysis of
this BCG sample, supplemented with galaxies from the next slices of the
survey, is underway.

\section{Conclusions}

We conducted follow-up spectroscopy within the third declination slice
of candidates from the Hamburg/SAO Survey for ELGs.
Summarizing the results presented, the analysis of the content of various
types of objects, and the discussion above, we draw the following
conclusions:

\begin{itemize}

\item The intended methods to detect ELG candidates on the plates of the
      Hamburg Quasar Survey give a reasonably high detection rate of
      emission-line objects ($\sim$~75~\%) (85 objects of 113 observed
      in this third part).

\item Besides ELGs, we found also 4 new quasars, all with Ly$\alpha$
      in the wavelength region $4950-5100$~\AA\, (i.e with 3.07 $< z < $ 3.2)
      near the red boundary of the IIIa-J photoplates.

\item The high fraction of BCG/H{\sc ii} galaxies among all newly
	discovered ELGs (about 68~\% in this paper) is in line with our main
	goal --- to pick up a deep BCG sample in the sky region under
	analysis.

\end{itemize}

\begin{acknowledgements}

This work was supported by the grant of the Deutsche
Forschungsgemeinschaft No.~436 RUS~17/77/94. U.A.V. is grateful
to the staff of the Hamburg Observatory for their hospitality and kind
assistance.  Y.I.I. thanks the staff of the National Optical Astronomy
Observatories for their kind hospitality.
Support by the INTAS grant No.~96-0500 was crucial to
proceed with the Hamburg/SAO survey declination band centered on
+37.5$^{\circ}$.  SAO authors appreciate the partial financial support
from the Russian Foundation for Basic Research grant No.~96-02-16398
and from the Russian Center of Cosmoparticle Physics ``Cosmion''.
The authors acknowledge the use of the NASA/IPAC Extragalactic
Database (NED).

\end{acknowledgements}

\clearpage




\scriptsize

\begin{table*}[h]

\begin{center}
\caption{\label{Tab2} Coordinates, Velocities and Magnitudes of
Emission--Line Galaxies}

\begin{tabular}{rlllrrccl} \hline \\[-0.35cm]
\multicolumn{1}{c}{\#}               &
\multicolumn{1}{c}{Name}             &
\multicolumn{1}{c}{$\alpha\,(1950)$} &
\multicolumn{1}{c}{$\delta\,(1950)$} &
\multicolumn{1}{c}{$v_{0}^{\,a}$}    &
\multicolumn{1}{r}{$m_{\rm B}$}          &
\multicolumn{1}{c}{$M_{\rm B}^{\;\;b}$}  &
\multicolumn{1}{c}{Type}             &
\multicolumn{1}{l}{Other Names from NED}\\
&
\multicolumn{1}{c}{ (1) } &
\multicolumn{1}{c}{ (2) } &
\multicolumn{1}{c}{ (3) } &
\multicolumn{1}{c}{ (4) } &
\multicolumn{1}{c}{ (5) } &
\multicolumn{1}{c}{ (6) } &
\multicolumn{1}{c}{ (7) } &
\multicolumn{1}{c}{ (8) } \\
\hline
\qq& HS 0940+3737* &09$^{\rm h}$40$^{\rm m}$10\fs7&37$^\circ$37$'$14$''$& 15655 $\pm$107 & 16.1 &--20.5 & Sy & IRAS F09401+3737 \\
\qq& HS 0948+3723  &09 48 19.0 &37 23 05 &  4806 $\pm$  30 &+17.8 &--16.2       & NON   & \\
\qq& HS 0954+3612  &09 54 01.1 &36 12 12 & 13602 $\pm$  24 & 17.3 &--19.0       & BCG?   & \\
\qq& HS 0956+3847  &09 56 51.0 &38 47 54 & 15785 $\pm$  21 & 16.7 &--19.9       & BCG   & \\
\qq& HS 1012+3655  &10 12 49.0 &36 55 31 & 10097 $\pm$  12 & 18.3 &--17.4       & BCG   & \\
\qq& HS 1015+3737* &10 15 23.3 &37 37 44 & 14533 $\pm$  66 & 16.4 &--20.0       & Sy & IRAS F10154+3737 ? \\
\qq& HS 1018+3619* &10 18 11.8 &36 19 16 & 22244 $\pm$  65 & 17.3 &--20.0       & BCG & \\
\qq& HS 1024+3648* &10 24 56.4 &36 48 54 & 11184 $\pm$  68 & 16.6 &--19.3       & Sy  & FIRST J102749.9+363334 \\
\qq& HS 1053+3722* &10 53 43.0 &37 22 55 & 13011 $\pm$  61 & 16.7 &--19.5       & SBN?   & \\
\qq& HS 1058+3553  &10 58 12.5 &35 53 15 &  9366 $\pm$  12 & 17.0 &--18.5       & BCG   & \\  
\qq& HS 1103+3627* &11 03 15.9 &36 27 30 & 13832 $\pm$  67 & 15.5 &--20.8       & BCG   & CG 810, IRAS  \\
\qq& HS 1103+3758* &11 03 03.4 &37 58 24 & 13664 $\pm$  60 & 18.1 &--18.2       & SBN   & \\
\qq& HS 1106+3621  &11 06 35.6 &36 21 52 &  6526 $\pm$  60 & 18.4 &--16.3       & BCG   & \\
\qq& HS 1107+3524  &11 07 53.1 &35 24 36 &  8362 $\pm$  48 & 17.4 &--17.8       & BCG   & \\  
\qq& HS 1107+3637  &11 07 04.0 &36 37 16 &  8206 $\pm$  18 & 19.5 &--15.7       & NON   & \\  
\qq& HS 1107+3710* &11 07 31.5 &37 10 49 &  8750 $\pm$  84 & 17.0 &--18.3       & SBN   & \\
\qq& HS 1107+3712  &11 07 17.7 &37 12 21 &  9639 $\pm$  13 & 17.6 &--17.9       & BCG?  & \\
\qq& HS 1133+3852  &11 33 50.2 &38 52 37 &  6631 $\pm$  27 &+19.8 &--14.9       & BCG?   & \\ 
\qq& HS 1134+3805* &11 34 28.0 &38 05 10 &  9578 $\pm$  70 &+18.6 &--16.9       & BCG   & \\
\qq& HS 1139+3712* &11 39 01.3 &37 12 19 &  6523 $\pm$  60 & 17.0 &--17.7       & BCG   & \\
\qq& HS 1148+3827* &11 48 00.3 &38 27 00 & 10334 $\pm$  65 & 18.7 &--17.0       & BCG   & \\
\qq& HS 1158+3837  &11 58 21.5 &38 37 41 &  6898 $\pm$  15 & 17.9 &--16.9       & BCG   & CG 1493  \\
\qq& HS 1159+3940  &11 59 52.5 &39 40 59 &  6350 $\pm$  41 &+17.4 &--17.2       & NON   & \\
\qq& HS 1207+3957  &12 07 14.2 &39 57 57 & 11700 $\pm$  30 & 18.5 &--17.5       & BCG   & \\
\qq& HS 1209+3726  &12 09 20.2 &37 26 57 &  7273 $\pm$  54 & 16.9 &--18.0       & BCG?   & \\ 
\qq& HS 1210+3759* &12 10 13.4 &37 59 40 &  6875 $\pm$  84 &+17.2 &--17.6       & DANS?   & \\
\qq& HS 1214+3922  &12 14 30.6 &39 22 22 &  4438 $\pm$  12 & 18.2 &--15.5       & BCG   & \\
\qq& HS 1214+3801* &12 14 39.6 &38 01 32 &   276 $\pm$  67 &+16.5 &--11.3       & SA in SpG   & part of NGC 4244  \\ 
\qq& HS 1237+3900  &12 37 04.0 &39 00 00 & 11199 $\pm$  21 &+18.6 &--17.3       & BCG   & \\ 
\qq& HS 1248+3523  &12 48 53.0 &35 23 06 & 10025 $\pm$  15 & 17.4 &--18.2       & BCG   & \\
\qq& HS 1249+4021* &12 49 25.1 &40 21 04 &  9882 $\pm$  67 & 17.0 &--18.6       & SBN?   & \\
\qq& HS 1254+3740  &12 54 55.2 &37 40 46 &  7635 $\pm$  15 & 18.0 &--17.0       & BCG   & \\
\qq& HS 1257+3658* &12 57 45.0 &36 58 05 &  8658 $\pm$  84 & 17.1 &--18.2       & BCG   & NGP9 F269-0458422  \\
\qq& HS 1302+3607  &13 02 49.9 &36 07 49 &  9561 $\pm$  27 & 17.1 &--18.4       & BCG   & CG 1080 \\ 
\qq& HS 1309+3701  &13 09 50.6 &37 01 54 &  7211 $\pm$  30 & 18.1 &--16.8       & BCG   & NGP9 F269-0480622 \\
\qq& HS 1311+3641* &13 11 54.7 &36 41 48 & 28685 $\pm$ 116 & 17.5 &--20.4       & SBN   & NGP9 F269-0606354 \\
\qq& HS 1323+3851* &13 23 43.1 &38 51 25 &  6580 $\pm$  61 & 17.5 &--17.2       & BCG   & \\
\qq& HS 1327+3731* &13 27 26.8 &37 31 28 &  4187 $\pm$  64 & 16.5 &--17.2       & BCG   & NGP9 F270-0319016 \\
\qq& HS 1331+3545* &13 31 59.8 &35 45 05 & 17661 $\pm$  93 & 18.1 &--18.8       & BCG   & \\
\qq& HS 1331+3657  &13 31 46.5 &36 57 21 & 18351 $\pm$  15 & 18.0 &--18.9       & BCG   & \\
\qq& HS 1333+3717* &13 33 34.6 &37 17 04 & 17008 $\pm$  33 & 18.0 &--18.8       & BCG   & HS 1333+3717\\
\qq& HS 1341+3700* &13 41 00.9 &37 00 01 &  5882 $\pm$  81 & 15.8 &--18.7       & SBN   & CG 1181=IRAS \\
\qq& HS 1347+3811  &13 47 02.3 &38 11 35 &  2828 $\pm$  32 & 18.8 &--14.1       & NON   & HS 1347+3811  \\
\hline
\end{tabular}
\end{center}
\end{table*}

\newpage

\begin{table*}[h]

\begin{center}
\flushleft {\bf Table 2.} (Continued)

\begin{tabular}{rlllrrccl} \\[-0.22cm] \hline \\[-0.35cm]
\multicolumn{1}{c}{\#}               &
\multicolumn{1}{c}{Name}             &
\multicolumn{1}{c}{$\alpha\,(1950)$} &
\multicolumn{1}{c}{$\delta\,(1950)$} &
\multicolumn{1}{c}{$v_{0}^{\,a}$}    &
\multicolumn{1}{r}{$m_{\rm B}$}          &
\multicolumn{1}{c}{$M_{\rm B}^{\;\;b}$}  &
\multicolumn{1}{c}{Type}             &
\multicolumn{1}{l}{Other Names from NED}\\
&
\multicolumn{1}{c}{ (1) } &
\multicolumn{1}{c}{ (2) } &
\multicolumn{1}{c}{ (3) } &
\multicolumn{1}{c}{ (4) } &
\multicolumn{1}{c}{ (5) } &
\multicolumn{1}{c}{ (6) } &
\multicolumn{1}{c}{ (7) } &
\multicolumn{1}{c}{ (8) } \\
\hline
\qq& HS 1402+3945  &14$^{\rm h}$02$^{\rm m}$55\fs0&39$^\circ$45$'$11$''$& 19654 $\pm$ 18& 16.6 &--20.5 & SBN & \\
\qq& HS 1414+3605* &14 14 50.8 &36 05 01 & 43503 $\pm$ 132 & 18.3 &--20.5 & Sy & \\
\qq& HS 1424+3836  &14 24 25.9 &38 36 26 &  6708 $\pm$  15 &+18.4 &--16.4 & BCG & CG 428=HS 1424+3836 \\
\qq& HS 1425+3847* &14 25 33.8 &38 47 17 &  6578 $\pm$  63 & 16.0 &--18.7 & SBN?  & CG 436 \\
\qq& HS 1428+3545  &14 28 23.3 &35 45 24 & 11858 $\pm$  27 &+18.5 &--17.5 & BCG  & \\
\qq& HS 1434+3644* &14 34 20.1 &36 44 26 &  9385 $\pm$  88 &+18.1 &--17.4 & BCG & \\
\qq& HS 1437+4002  &14 37 58.5 &40 02 29 & 11068 $\pm$  31 &+19.2 &--16.7 & BCG?& \\
\qq& HS 1447+3636* &14 47 48.0 &36 36 03 &  1861 $\pm$  67 & 17.7 &--14.3 & BCG & \\
\qq& HS 1449+3647  &14 49 02.3 &36 47 24 & 10086 $\pm$  15 &+18.6 &--17.1 & BCG & \\  
\qq& HS 1457+3652  &14 57 59.6 &36 52 39 & 11768 $\pm$  18 &+19.8 &--16.2 & BCG & \\
\qq& HS 1458+3726* &14 58 32.4 &37 26 02 & 32205 $\pm$ 306 &+20.7 &--17.5 & NON & \\
\qq& HS 1513+3612* &15 13 04.1 &36 12 05 & 18466 $\pm$  67 &+17.8 &--19.2 & BCG  & \\
\qq& HS 1519+4007* &15 19 30.0 &40 07 39 &  2625 $\pm$  60 &+16.5 &--16.2 & BCG  & CG 691 \\
\qq& HS 1520+3717  &15 20 30.6 &37 17 56 & 11144 $\pm$  78 &+18.1 &--17.8 & NON & \\
\qq& HS 1524+3536* &15 24 48.8 &35 36 39 &  8770 $\pm$  12 & 18.0 &--17.3 & BCG  & CG 720  \\  
\qq& HS 1526+3729* &15 26 18.7 &37 29 36 &  9792 $\pm$  64 &+16.5 &--19.1 & Sy & IRAS F15263+3729 \\
\qq& HS 1526+3634  &15 26 27.2 &36 34 29 & 18825 $\pm$  20 &+19.1 &--17.9 & BCG & \\  
\qq& HS 1527+3912  &15 27 46.2 &39 12 30 & 23004 $\pm$  38 &+15.7 &--21.7 & SBN & CG 734 \\
\qq& HS 1529+4003* &15 29 45.1 &40 03 21 &  8379 $\pm$  60 &+16.7 &--18.5 & BCG? & CG 749 \\
\qq& HS 1530+3617  &15 30 57.5 &36 17 40 &  4666 $\pm$  24 &+17.0 &--17.0 & DANS? & \\
\qq& HS 1535+3747* &15 35 40.6 &37 47 47 &  7660 $\pm$  63 &+19.1 &--16.0 & BCG & \\
\qq& HS 1536+3958  &15 36 52.3 &39 58 10 &  4367 $\pm$  18 & 18.0 &--15.8 & BCG & \\
\qq& HS 1543+3830  &15 43 44.8 &38 30 30 &  7862 $\pm$  18 & 19.2 &--15.9 & BCG & \\
\qq& HS 1557+3549* &15 57 35.8 &35 49 45 & 12852 $\pm$  97 & 16.8 &--19.4 & BCG? & \\
\qq& HS 1558+3543  &15 58 27.3 &35 43 15 & 20377 $\pm$  84 & 18.3 &  18.9 & NON  & \\
\qq& HS 1558+3636* &15 58 34.1 &36 36 34 & 12495 $\pm$  62 & 17.8 &--18.3 & BCG & \\
\qq& HS 1608+3654  &16 08 31.5 &36 54 42 &  9587 $\pm$  18 & 17.3 &--18.2 & NON & \\
\qq& HS 1613+3701  &16 13 50.6 &37 01 13 &  8844 $\pm$  21 & 17.1 &--18.3 & BCG & \\
\qq& HS 1615+3521* &16 15 48.4 &35 21 07 &  8507 $\pm$  70 & 17.5 &--17.8 & BCG & \\
\qq& HS 1619+3557* &16 19 11.3 &35 57 35 & 15092 $\pm$  42 & 17.4 &--19.1 & BCG  & \\
\qq& HS 1624+3618  &16 24 22.3 &36 18 25 &  9484 $\pm$  12 & 19.0 &--16.5 & BCG & \\
\qq& HS 1638+3938  &16 38 01.8 &39 38 17 &  8969 $\pm$  15 & 18.2 &--17.2 & BCG & \\
\qq& HS 1643+4015  &16 43 39.0 &40 15 07 & 11617 $\pm$  18 & 17.4 &--18.6 & BCG & \\  
\qq& HS 1645+3551* &16 45 51.9 &35 51 57 &  6720 $\pm$  61 & 17.7 &--17.1 & BCG & \\
\qq& HS 1653+3634  &16 53 07.2 &36 34 55 &   845 $\pm$  24 &+16.0 &--14.3 & BCG & \\
\qq& HS 1656+3927  &16 56 34.7 &39 27 58 & 10290 $\pm$  42 & 17.7 &--18.0 & Sy & FIRST J165815.4+392329 \\  
\qq& HS 1711+3756* &17 11 41.5 &37 56 12 &  8273 $\pm$  59 & 17.6 &--17.6 & BCG & \\
\qq& HS 1721+3615  &17 21 47.5 &36 15 43 & 15596 $\pm$  15 & 18.1 &--18.5 & BCG & \\
\hline
\multicolumn{8}{l}{ $^a$ Heliocentric velocities.} \\
\multicolumn{8}{l}{ $^b$ Absolute magnitudes are not corrected for the galactic extinction.} \\
\multicolumn{8}{l}{ * Objects observed with the 2.2\,m Calar Alto telescope.} \\
\multicolumn{8}{l}{ + APM magnitude. } \\
\end{tabular}
\end{center}
\end{table*}




\setcounter{qub}{0}

\begin{table*}[h]

\begin{center}
\caption{\label{Tab3} Coordinates, Redshifts and Magnitudes of QSO }

\begin{tabular}{lccrcl} \hline \\[-0.35cm]
\MC{1}{c}{Name}             &
\MC{1}{c}{$\alpha\,(1950)$} &
\MC{1}{c}{$\delta\,(1950)$} &
\MC{1}{c}{$z^{\,a}$}        &
\MC{1}{c}{$m_{\rm B}$}          &
\MC{1}{c}{Detected Emission Lines}\\
\MC{1}{c}{ (1) } &
\MC{1}{c}{ (2) } &
\MC{1}{c}{ (3) } &
\MC{1}{c}{ (4) } &
\MC{1}{c}{ (5) } &
\MC{1}{c}{ (6) }
\\ \hline
HS 1003+3719  &10$^{\rm h}$03$^{\rm m}$12\fs5&37$^\circ$19 $'$51$''$& 3.204 &~17.7     &Ly$\alpha$ 1216 \AA, Si{\sc iv}/O{\sc iv}] 1400 \AA, C{\sc iv} 1549 \AA  \\
HS 1024+3558  &10  \,24\,\,21.7    &35      \,58\,\,36    & 3.106 &~16.9     &Ly$\alpha$ 1216 \AA, Si{\sc iv}/O{\sc iv}] 1400 \AA, C{\sc iv} 1549 \AA  \\
HS 1152+3700  &11  \,52\,\,50.0    &37      \,00\,\,20    & 3.069 &~20.0     &Ly$\alpha$ 1216 \AA                      \\
HS 1649+3905  &16  \,49\,\,02.6    &39      \,05\,\,46    & 3.166 &~18.3     &Ly$\alpha$ 1216 \AA, C{\sc iv} 1549 \AA  \\
\hline \\[-0.2cm]
\multicolumn{6}{l}{ $^a$ Observed redshift. } \\
\end{tabular}
\end{center}
\end{table*}




\scriptsize
\setcounter{qub}{0}

\begin{table*}[h]

\begin{center}
\caption{\label{Tab4} Parameters of Emission Lines of Galaxies}

\begin{tabular}{rlrcccccrrr} \hline \\[-0.35cm]
\multicolumn{1}{c}{\#}                             &
\multicolumn{1}{c}{ Name}                          &
\multicolumn{1}{c}{ $F$(H$\beta$)$^a$ }              &
\multicolumn{1}{c}{ $\frac{F(\lambda 3727)}{F(\rm H\beta)}$} &
\multicolumn{1}{c}{ $\frac{F(\lambda 5007)}{F(\rm H\beta)}$} &
\multicolumn{1}{c}{ $\frac{F(\rm H\alpha)}{F(\rm H\beta)}$} &
\multicolumn{1}{c}{ $\frac{F(\lambda 6583)}{F(\rm H\alpha)}$}&
\multicolumn{1}{c}{ $\frac{F([\rm SII])}{F(\rm H\alpha)}$} &
\multicolumn{1}{c}{ $W_{\lambda 3727}$(\AA)}       &
\multicolumn{1}{c}{ $W_{\rm H\beta}$(\AA)}             &
\multicolumn{1}{c}{ $W_{\lambda 5007}$(\AA)}       \\
&
\multicolumn{1}{c}{ (1) } &
\multicolumn{1}{c}{ (2) } &
\multicolumn{1}{c}{ (3) } &
\multicolumn{1}{c}{ (4) } &
\multicolumn{1}{c}{ (5) } &
\multicolumn{1}{c}{ (6) } &
\multicolumn{1}{c}{ (7) } &
\multicolumn{1}{c}{ (8) } &
\multicolumn{1}{c}{ (9) } &
\multicolumn{1}{c}{ (10) } \\
\hline
\qq& HS 0940+3737$^*$&    48:&    -- & 14.00 & 22.17 &  0.64 &  0.21  &   -- &    5 &   73 \\
\qq& HS 0948+3723    &   +20 &    -- &    -- &    -- &    -- &    --  &   -- &  +76 &   58 \\
\qq& HS 0954+3612    &    84 &  1.71 &  3.21 &  3.16 &  0.11 &  0.30  &   55 &   21 &   72 \\
\qq& HS 0956+3847    &    68 &  3.02 &  4.06 &  2.93 &  0.09 &  0.06: &  295 &   59 &  225 \\
\qq& HS 1012+3655    &    49 &  1.50 &  4.53 &  3.42 &  0.11 &  0.16: &   67 &   55 &  302 \\
\qq& HS 1015+3737$^*$&  +133:&    -- &    -- &    -- &  0.58 &  0.73  &   -- &  +27 &   94 \\
\qq& HS 1018+3619$^*$&    68:&  3.78 &  4.15 &  3.38 &  0.06:&  0.17  &   98 &   47 &  212 \\
\qq& HS 1024+3648$^*$&    40:&    -- &  4.20 &  8.61 &  0.59 &  0.33  &   -- &    8 &   37 \\
\qq& HS 1053+3722$^*$&    39:&    -- &  3.35 &  5.76 &  0.23 &  0.31  &   -- &   15 &   53 \\
\qq& HS 1058+3553    &    36 &  4.14 &  5.72 &  5.47 &  0.12 &  0.29  &   43 &   11 &   68 \\
\qq& HS 1103+3627$^*$&   109:&    -- &  2.92 &  5.47 &  0.25 &  0.13  &   -- &   12 &   35 \\
\qq& HS 1103+3758$^*$&    25:&    -- &  1.51 &  1.96 &  0.10:&  0.39  &   -- &   20 &   33 \\
\qq& HS 1106+3621    &    14 &  1.04 & 10.51 &  4.92 &    -- &    --  &   58 &   17 &  197 \\
\qq& HS 1107+3524    &    13 &  5.69 &  4.04 &  4.37 &    -- &  0.39: &   33 &    7 &   28 \\
\qq& HS 1107+3637    &   +45 &    -- &    -- &    -- &    -- &    --  &   -- & +316 &  144 \\
\qq& HS 1107+3710$^*$&  +117:&    -- &    -- &    -- &  0.65 &  0.23  &   -- &  +44 &   -- \\
\qq& HS 1107+3712    &    27 &  3.28 &  4.64 &  4.00 &    -- &    --  &   23 &   12 &   60 \\
\qq& HS 1133+3852    &    15 &  4.04 & 13.92 &  5.04 &  0.07 &    --  &  146 &   22 &  332 \\
\qq& HS 1134+3805$^*$&    77:&  2.81 &  4.20 &  2.65 &    -- &  0.18  &  440 &   65 &  253 \\
\qq& HS 1139+3712$^*$&    32:&    -- &  3.45 &  3.75 &  0.12 &  0.18  &   -- &   71 &  259 \\
\qq& HS 1148+3827$^*$&    49:&  0.72 &  5.70 &  4.05 &  0.05:&  0.09  &  147 &   79 &  469 \\
\qq& HS 1158+3837    &    93 &  2.37 &  3.67 &  3.37 &  0.05 &  0.08  &   75 &   27 &  105 \\
\qq& HS 1159+3940    &   +31 &    -- &    -- &    -- &    -- &    --  &   -- &  +74 &   22 \\
\qq& HS 1207+3957    &    79 &  1.47 &  3.86 &  2.29 &    -- &  0.07: &   69 &   42 &  165 \\
\qq& HS 1209+3726    &     9 &    -- & 11.65 &  9.18 &    -- &    --  &   -- &    4 &   45 \\
\qq& HS 1210+3759$^*$&   +32:&    -- &    -- &    -- &  0.15 &    --  &   -- &  +20 &    9 \\
\qq& HS 1214+3922    &   621 &  0.85 &  4.73 &  2.79 &    -- &  0.03: &  136 &  105 &  522 \\
\qq& HS 1214+3801$^*$&   683:&  0.49 &  5.44 &  2.74 &  0.05:&  0.07  &  101 &  313 & 1557 \\
\qq& HS 1237+3900    &    28 &  0.68 &  3.79 &  2.52 &  0.20:&    --  &   39 &   36 &  138 \\
\qq& HS 1248+3523    &    82 &  1.67 &  3.29 &  3.39 &    -- &  0.06  &   56 &   38 &  136 \\
\qq& HS 1249+4021$^*$&    28:&    -- &  3.25 &  5.66 &  0.29 &  0.51  &   -- &    7 &   24 \\
\qq& HS 1254+3740    &    75 &  1.06 &  6.33 &  3.57 &    -- &    --  &   46 &   53 &  336 \\
\qq& HS 1257+3658$^*$&    15:&    -- &  6.21 &  2.94 &  0.10 &  0.17  &   -- &   36 &  255 \\
\qq& HS 1302+3607    &    88 &  1.70 &  4.05 &  3.25 &  0.14 &  0.17: &   19 &   21 &   88 \\
\qq& HS 1309+3701    &    26 &  2.15 &  4.77 &  5.04 &    -- &  0.09: &   43 &   25 &  128 \\
\qq& HS 1311+3641$^*$&   +49:&    -- &    -- &    -- &  0.32 &  0.43  &   -- &  +36 &   -- \\
\qq& HS 1323+3851$^*$&    15:&    -- &  4.28 &  4.27 &  0.03 &  0.11  &   -- &   13 &   56 \\
\qq& HS 1327+3731$^*$&    36:&    -- &  5.90 &  7.26 &   --  &  0.17  &   -- &    9 &   53 \\
\qq& HS 1331+3545$^*$&    74:&  3.27 &  4.64 &  2.89 &  0.02:&  0.16  &  114 &   62 &  295 \\
\qq& HS 1331+3657    &    21 &  2.13 &  6.43 &  3.39 &    -- &  0.13: &  328 &  140 &  962 \\
\qq& HS 1333+3717$^*$&   193:&  2.30 &  4.72 &  3.27 &  0.09 &  0.13  &  161 &  149 &  648 \\
\qq& HS 1341+3700$^*$&    54:&    -- &  1.93 &  4.51 &  0.30 &  0.39  &   -- &   15 &   31 \\
\qq& HS 1347+3811    &   +28 &    -- &    -- &    -- &    -- &    --  &  249 &  +51 &   85 \\
\hline  \\[--0.2cm]
\end{tabular}
\end{center}
\end{table*}

\newpage

\begin{table*}[h]

\begin{center}
\flushleft {\bf Table 4.} (Continued)
\begin{tabular}{rlrcccccrrr} \hline \\[-0.35cm]
\multicolumn{1}{c}{\#}                             &
\multicolumn{1}{c}{ Name}                          &
\multicolumn{1}{c}{ $F$(H$\beta$)$^a$ }              &
\multicolumn{1}{c}{ $\frac{F(\lambda 3727)}{F(\rm H\beta)}$} &
\multicolumn{1}{c}{ $\frac{F(\lambda 5007)}{F(\rm H\beta)}$} &
\multicolumn{1}{c}{ $\frac{F(\rm H\alpha)}{F(\rm H\beta)}$} &
\multicolumn{1}{c}{ $\frac{F(\lambda 6583)}{F(\rm H\alpha)}$}&
\multicolumn{1}{c}{ $\frac{F([\rm SII])}{F(\rm H\alpha)}$} &
\multicolumn{1}{c}{ $W_{\lambda 3727}$(\AA)}       &
\multicolumn{1}{c}{ $W_{\rm H\beta}$(\AA)}             &
\multicolumn{1}{c}{ $W_{\lambda 5007}$(\AA)}       \\
&
\multicolumn{1}{c}{ (1) } &
\multicolumn{1}{c}{ (2) } &
\multicolumn{1}{c}{ (3) } &
\multicolumn{1}{c}{ (4) } &
\multicolumn{1}{c}{ (5) } &
\multicolumn{1}{c}{ (6) } &
\multicolumn{1}{c}{ (7) } &
\multicolumn{1}{c}{ (8) } &
\multicolumn{1}{c}{ (9) } &
\multicolumn{1}{c}{ (10) } \\
\hline
\qq& HS 1402+3945    &    19 &  2.05 &  1.53 &  9.37 &  0.48 &  0.30  &   13 &    6 &   11 \\
\qq& HS 1414+3605$^*$&    +8:&    -- &    -- &    -- &  1.76 &    --  &   -- &  +40 &   61 \\
\qq& HS 1424+3836    &    51 &  0.92 &  7.01 &  3.37 &    -- &  0.05: &  184 &   73 &  556 \\
\qq& HS 1425+3847$^*$&    59:&    -- &  3.17 &  4.36 &  0.13 &  0.23  &   -- &   16 &   50 \\
\qq& HS 1428+3545    &    47 &  2.26 &  6.20 &  3.87 &    -- &    --  &   94 &   41 &  271 \\
\qq& HS 1434+3644$^*$&    27:&    -- &  5.93 &  3.02 &  0.02:&  0.13  &   -- &   48 &  290 \\
\qq& HS 1437+4002    &    12 &    -- &  4.60 &  3.00 &    -- &    --  &   -- &   30 &  128 \\
\qq& HS 1447+3636$^*$&    61:&    -- &  4.59 &  1.95 &  0.06 &  0.10  &   -- &   79 &  408 \\
\qq& HS 1449+3647    &    83 &  1.10 &  5.51 &  3.24 &    -- &  0.11  &   33 &   84 &  414 \\
\qq& HS 1457+3652    &   100 &    -- &  3.82 &  1.66 &    -- &    --  &   -- &   67 &  279 \\
\qq& HS 1458+3726$^*$&   +12:&    -- &    -- &    -- &    -- &    --  &   -- &  +11 &   -- \\
\qq& HS 1513+3612$^*$&    45:&  2.20 &  5.47 &  5.55 &    -- &  0.16  &   89 &   53 &  273 \\
\qq& HS 1519+4007$^*$&    72:&  0.66 &  5.77 &  2.90 &    -- &  0.10  &   19 &   34 &  200 \\
\qq& HS 1520+3717    &    11 &    -- &  6.79 &  4.64 &    -- &    --  &   -- &    6 &   47 \\
\qq& HS 1524+3536$^*$&    86:&  1.54 &  5.41 &  2.70 &  0.08 &  0.09  &   84 &   74 &  320 \\
\qq& HS 1526+3634    &   110 &  1.14 &  4.65 &  2.36 &    -- &  0.23  &  136 &   77 &  385 \\
\qq& HS 1526+3729$^*$&   +29:&    -- &    -- &    -- &  1.52 &  0.94  &   -- &   +6 &   10 \\
\qq& HS 1527+3912    &    36 &  3.39 &  1.17 &  3.77 &  0.17 &  0.22  &   28 &   10 &   13 \\
\qq& HS 1529+4003$^*$&    30:&    -- &  3.03 &  5.19 &  0.24 &  0.20  &   -- &    9 &   29 \\
\qq& HS 1530+3617    &    12 &    -- &  4.29 &  3.38 &    -- &    --  &   -- &    4 &   19 \\
\qq& HS 1535+3747$^*$&    14:&    -- &  5.20 &  2.95 &    -- &    --  &   -- &   33 &  178 \\
\qq& HS 1536+3958    &    90 &    -- &  4.52 &  4.01 &    -- &  0.27  &   -- &   24 &  114 \\
\qq& HS 1558+3543    &   +42 &    -- &    -- &    -- &    -- &    --  &   49 &  +73 &   17 \\
\qq& HS 1543+3830    &   115 &  2.33 &  4.21 &  3.01 &  0.09 &  0.08: &   78 &   27 &  123 \\
\qq& HS 1557+3549$^*$&    13:&    -- &  3.72 &  3.98 &    -- &  0.24  &   -- &   14 &   54 \\
\qq& HS 1558+3636$^*$&    49:&    -- &  2.64 &  3.63 &  0.04:&  0.25  &   -- &   24 &   68 \\
\qq& HS 1608+3654    &  +105 &    -- &    -- &    -- &    -- &  0.22  &   -- &  +60 &   80 \\
\qq& HS 1613+3701    &    49 &  3.13 &  4.10 &  5.61 &  0.10 &  0.18  &   48 &   14 &   62 \\
\qq& HS 1615+3521$^*$&     7:&    -- &  5.30 &  5.72 &  0.06 &  0.16  &   -- &    5 &   26 \\
\qq& HS 1619+3557$^*$&    22:&    -- &  4.22 &  4.27 &  0.06 &  0.22  &   -- &   21 &   89 \\
\qq& HS 1624+3618    &    31 &  1.07 &  5.96 &    -- &    -- &    --  &   67 &   70 &  390 \\
\qq& HS 1638+3938    &    87 &  1.58 &  6.13 &  3.74 &  0.04:&  0.19  &   77 &   62 &  367 \\
\qq& HS 1643+4015    &    87 &  2.45 &  3.32 &  3.46 &  0.08 &  0.17  &   49 &   23 &   81 \\
\qq& HS 1645+3551$^*$&    16:&    -- &  5.34 &  3.41 &  0.05:&  0.09  &   -- &   23 &  133 \\
\qq& HS 1653+3634    &   346 &  1.08 &  5.16 &  2.37 &  0.03 &  0.11  &  124 &  102 &  561 \\
\qq& HS 1656+3927    &    31:&    -- & 21.41 &  3.33 &  1.59 &  0.58  &   -- &    4:&   81 \\
\qq& HS 1711+3756$^*$&    89:&  3.46 &  4.38 &  3.98 &    -- &  0.20  &  108 &   33 &  146 \\
\qq& HS 1721+3615    &    57 &  1.72 &  7.25 &  4.28 &  0.10 &  0.16  &   54 &   45 &  363 \\
\hline  \\[--0.2cm]
\multicolumn{10}{l}{ * Objects observed under poor photometric conditions. } \\
\multicolumn{10}{l}{ $^a$ Flux in 10$^{-16}$ ergs s$^{-1}$ cm$^{-2}$ \AA$^{-1}$ .} \\
\multicolumn{10}{l}{ $^+$ Parameters for H$\alpha$ emission line.} \\
\multicolumn{10}{l}{ : Parameters with less confident values. }
\end{tabular}
\end{center}
\end{table*}




\begin{table*}[h]

\begin{center}
\caption{\label{Tab5} Galaxies with Nondetected Emission Lines }

\begin{tabular}{lllrccl} \hline \\[-0.35cm]
\multicolumn{1}{c}{Name} &
\multicolumn{1}{c}{$\alpha\,(1950)$} &
\multicolumn{1}{c}{$\delta\,(1950)$} &
\multicolumn{1}{c}{$v^{\,a}_{0}$} &
\multicolumn{1}{c}{$m_{\rm B}$} &
\multicolumn{1}{r}{$M_{\rm B}^{\,b}$} &
\multicolumn{1}{l}{Absorption lines} \\
\MC{1}{c}{ (1) } &
\MC{1}{c}{ (2) } &
\MC{1}{c}{ (3) } &
\MC{1}{c}{ (4) } &
\MC{1}{c}{ (5) } &
\MC{1}{c}{ (6) } &
\MC{1}{c}{ (7) }
\\ \hline
HS 1639+3535  &$16^{\rm h} 39^{\rm m} 25\fs7$ &$35^{\circ} 35{'} 49{''}$ & 21285: &~18.0 &--19.3 & G$_{\rm band}$, NaD, H$\alpha$       \\
HS 1711+3619  &17 11 05.4       &36 19 50       & 5696   &+19.2 &--15.2 & H$\beta$, Mgb, NaD, H$\alpha$    \\
\hline \\[-0.2cm]
\multicolumn{7}{l}{ $^a$ Heliocentric velocities. } \\
\multicolumn{7}{l}{ $^b$ Absolute magnitudes are not corrected for the galactic extinction. }\\
\multicolumn{7}{l}{ $^+$ APM magnitude.} \\
\multicolumn{7}{l}{ : less confident values.}

\end{tabular}
\end{center}

\end{table*}




\setcounter{qub}{0}

\begin{table*}[t]
\begin{center}
\caption{\label{Tab6}Stars}
\begin{tabular}{rcccrcl} \hline \\[-0.35cm]
\multicolumn{1}{c}{\#}             &
\multicolumn{1}{c}{Name}             &
\multicolumn{1}{c}{$\alpha\,(1950)$} &
\multicolumn{1}{c}{$\delta\,(1950)$} &
\multicolumn{1}{r}{$m_{\rm B}$}          &
\multicolumn{1}{l}{Type}             &
\multicolumn{1}{c}{Spectral features} \\
&
\MC{1}{c}{ (1) } &
\MC{1}{c}{ (2) } &
\MC{1}{c}{ (3) } &
\MC{1}{c}{ (4) } &
\MC{1}{c}{ (5) } &
\MC{1}{c}{ (6) }
\\ \hline
\qq &HS 1002+3900   &$10^{\rm h}02^{\rm m}47\fs5$ &$39^{\circ}00{'}24{''}$&16.9  &  A-F& CaK, CaH, G$_{\rm band}$, H$\beta$, NaD?, H$\alpha$ \\
\qq &HS 1202+4023   &12 02 04.0 &40 23 23 &+19.0 & M & \\
\qq &HS 1207+3818   &12 07 51.5 &38 18 18 &+19.5 & A  & H$\delta$, H$\gamma$, H$\beta$, Mgb, H$\alpha$\\
\qq &HS 1207+4007   &12 07 44.9 &40 07 52 &+20.0 & M & \\
\qq &HS 1218+3857   &12 18 15.0 &38 57 12 &+19.9 & K  & TiI,II4533, Mgb, NaD, TiO 6156-6180, H$\alpha$\\
\qq &HS 1334+3521   &13 34 37.6 &35 21 39 &+19.3 & M & \\
\qq &HS 1513+3616   &15 13 14.5 &36 16 58 &17.9  & F-G & CaK, CaH, G$_{\rm band}$, H$\beta$, NaD, H$\alpha$ \\
\qq &HS 1514+3908   &15 14 23.3 &39 08 43 &+18.5 & F-G & CaK, CaH, G$_{\rm band}$,  Mgb\\
\qq &HS 1518+3600   &15 18 55.1 &36 00 44 &+18.7 & F-G & H$\beta$, Mgb, NaD, H$\alpha$\\
\qq &HS 1609+3909   &16 09 57.3 &39 09 13 &18.5  & F-G & CaK,  H$\beta$, Mgb, NaD, H$\alpha$\\
\qq &HS 1652+3851   &16 52 15.1 &38 51 39 &19.1  &     & MgFe 4703, Mgb, NaD\\
\qq &HS 1726+3514   &17 26 47.6 &35 14 42 &17.4  & A-F & CaK, CaH, G$_{\rm band}$, H$\beta$, Mgb, NaD, H$\alpha$\\
\qq &HS 1730+3607   &17 30 36.1 &36 07 22 &18.3  & F-G & CaK, CaH, H$\beta$, H$\alpha$\\
\hline \\[--0.2cm]
\multicolumn{7}{l}{ $^+$ APM magnitude.} \\
\end{tabular}
\end{center}
\end{table*}


\clearpage




\begin{thebibliography}{99}

\bibitem[1981]{Baldwin81}
 Baldwin J.A., Phillips M.M., Terlevich R., 1981, PASP 93, 5
\bibitem[1994]{Coziol94}
 Coziol R., Demers S., Pena M., Barneoud R., 1994, AJ 108, 405C
\bibitem[1993]{Coziol93}
 Coziol R., Demers S., Pena M., Torres-Peimbert S., Fontain G., Wesemael F.,
  Lamontagne R., 1993, MNRAS 261, 170
\bibitem[1994]{Engels94}
 Engels D., Cordis L., K\"{o}hler T., 1994, Proc. IAU Symp. 161, eds.
 H.T. MacGillivray, Kluwer, Dordrecht, p. 317
\bibitem[1997]{Gallego97}
 Gallego J., Zamorano J., Rego M., Vitores A.G., 1997, ApJ 475, 502
\bibitem[1989]{Grosbol89}
 Grosb{\o}l P., Reviews in modern astronomy, 1989, ed. G.Clark, 2, 242.
\bibitem[1995]{Hagen95}
 Hagen H.-J., Groote D., Engels, D., Reimers D., 1995, A\&AS 111, 195
\bibitem[1989]{Huchtmeier89} Huchtmeier W., Richter O., A General
Catalog of HI Observations of Galaxies, 1989, New York, Springer-Verlag
\bibitem[1993a]{Izotov93a}
 Izotov Y.I., Guseva N.G., Lipovetsky V.A. et al., 1993a,
 Astronomy \& Astrophysics Transactions 3, 179
\bibitem[1993b]{Izotov93b}
 Izotov Y.I., Lipovetsky V.A., Guseva N.G., 1993b, in:
 ``The Feedback of Chemical Evolution on  the Stellar Content of Galaxies'',
 eds. D.Alloin \& G.Stasinska, 127
\bibitem[1997]{Izotov97}
 Izotov Yu.I., Thuan T.X., Lipovetsky V.A., 1997, ApJS 108, 1
\bibitem[2000]{Kniazev2000} Kniazev A.Yu., Pustilnik S.A., Ugryumov A.V.,
  Kniazeva T.F. 2000, Astronomy Letters, v.26, issue 2 (in press)..
\bibitem[1983]{Markarian83}
 Markarian B.E., Lipovetsky V.A., Stepanian J.A., 1983, Afz 19, 29
\bibitem[1988]{Massey88}
 Massey P., Strobel K., Barnes J.V., Anderson E., 1988, ApJ 328, 315
\bibitem[1991]{Maza91}
 Maza J., Ruiz M.T., Gonz\'{a}lez L.E., Wischnjewsky M., Pe\~{n}a M.,
 1991, A\&AS 89, 389
\bibitem[1995]{Pesch95}
 Pesch P., Stephenson C.B., MacConnell D.J., 1995, ApJS 98, 41
\bibitem[1997]{Popescu97}
  Popescu C.C., Hopp U., Els\"asser H., 1997, A\&A 328, 756
\bibitem[1996]{Popescu96}
 Popescu C.C., Hopp U., Hagen H.-J., Els\"{a}sser H., 1996, A\&AS 116, 43
\bibitem[1998]{Popescu98}
 Popescu C.C., Hopp U., Hagen H.J., Els\"{a}sser H., 1998, A\&AS 133, 13
\bibitem[1999]{Pustilnik99}
 Pustilnik S.A., Engels D., Ugryumov A.V., Lipovetsky V.A., Hagen H.-J.,
 Kniazev A.Y., Izotov Y.I., Richter G., 1999, A\&AS 135, 299 ({\bf Paper~II})
\bibitem[1995]{Pustilnik95}
 Pustilnik S.A., Ugryumov A.V., Lipovetsky V.A., Thuan T.X., Guseva N.G., 1995,
 ApJ 443, 499
\bibitem[1989]{Salzer89}
 Salzer J.J., 1989, ApJ 347, 152
\bibitem[1988]{Salzer88}
 Salzer J.J., McAlpine G.M., 1988, AJ 96, 1192
\bibitem[1989]{Salzeretal89}
 Salzer J.J., MacAlpine G.M., Boroson T.A., 1989, ApJS 70, 479
\bibitem[1995]{Salzer95}
 Salzer J.J., Moody J.W., Rosenberg J.L., Gregory S.A., Newberry M.V.,
 1995, AJ 109, 2376
\bibitem[1994]{Stepanian94}
 Stepanian J.A., 1994, Proc. IAU Symp. 161, Kluwer, Dordrecht, eds.
 H.T. MacGillivray, p. 731
\bibitem[1998]{Surace98}
 Surace C., Comte G., 1998, A\&AS 133, 171
\bibitem[1991]{Terlevich91}
 Terlevich R., Melnick J., Masegosa J., Moles M., Copetti M.V.F., 1991,
 A\&AS 91, 285
\bibitem[1994]{Thuan94}
 Thuan T.X., Izotov Y.I., Lipovetsky V.A., Pustilnik S.A., 1994,
 Proc. ESO/OHP Workshop ``Dwarf Galaxies'', eds. Meylan \& Prugniel, p. 421
\bibitem[1999]{Thuan99}
 Thuan T.X., Lipovetsky V.A., Martin J.-M., Pustilnik S.A., 1999,
A\&AS, 139, 1 
\bibitem[1997]{Ugryumov97}
 Ugryumov A.V., 1997, Ph.D. Thesis, SAO RAS
\bibitem[1999]{Ugryumov99}
 Ugryumov A.V., Engels D., Lipovetsky V.A., Hagen H.-J., Hopp U.,
 Richter G.M., Pustilnik S.A., Kniazev A.Y., Izotov Y.I., Popescu C.C.,
 1999, A\&AS 135, 511 ({\bf Paper~I})
\bibitem[1998]{Ugryumov98}
 Ugryumov A.V., Pustilnik S.A., Lipovetsky V.A., Izotov Y.I., Richter G.M.,
1998, A\&AS 131, 295
\bibitem[1987]{Veilleux87}
 Veilleux S., Osterbrock D.E., 1987, ApJS 63, 295
\bibitem[1996]{Zamorano96}
 Zamorano J., Gallego J.G., Rego M., Vitores A. G., Alonso O.,
 1996, ApJS 105, 343
\bibitem[1994]{Zamorano94}
 Zamorano J., Rego M., Gallego J. et al., 1994, ApJS 95, 387

\end{thebibliography}
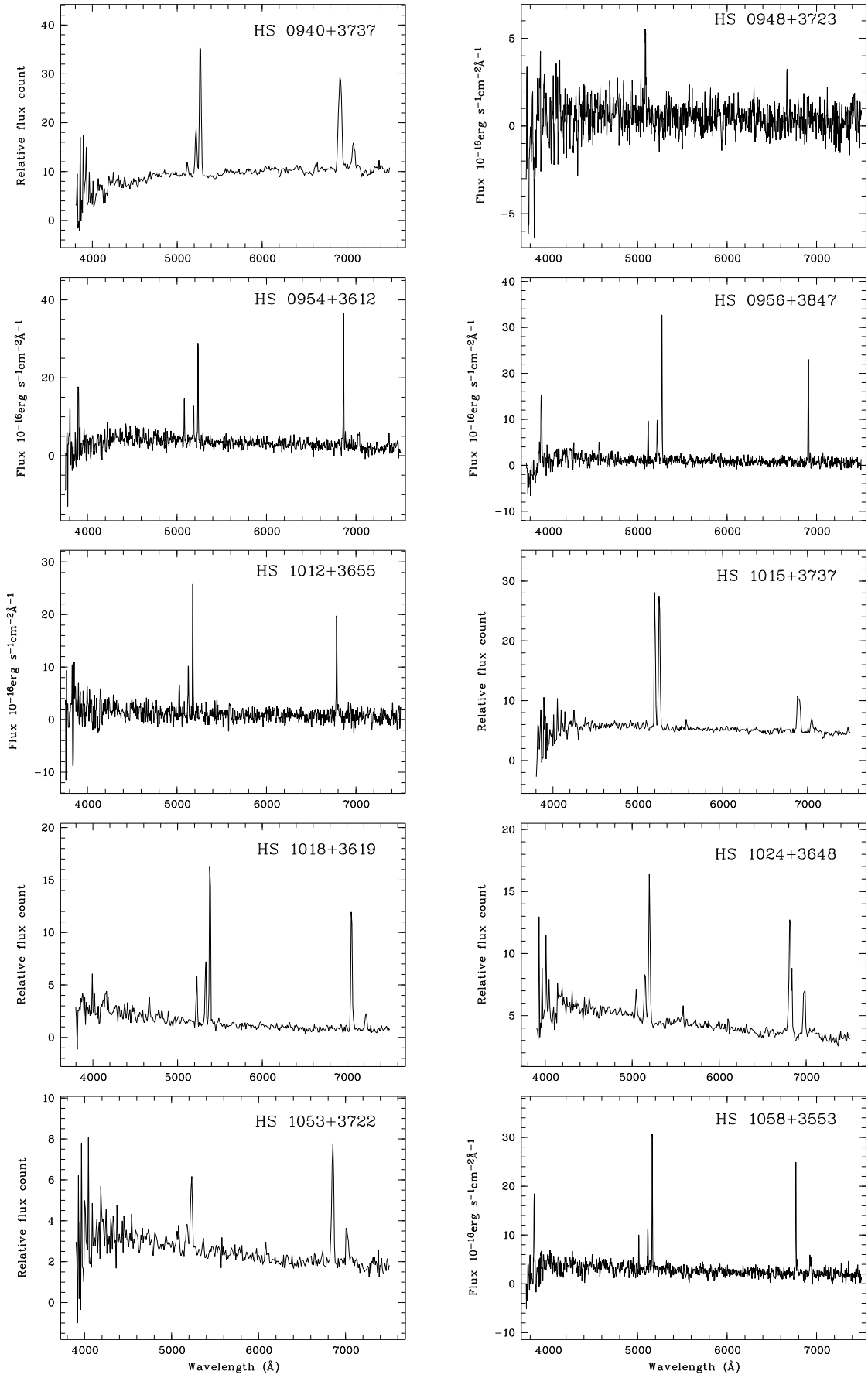
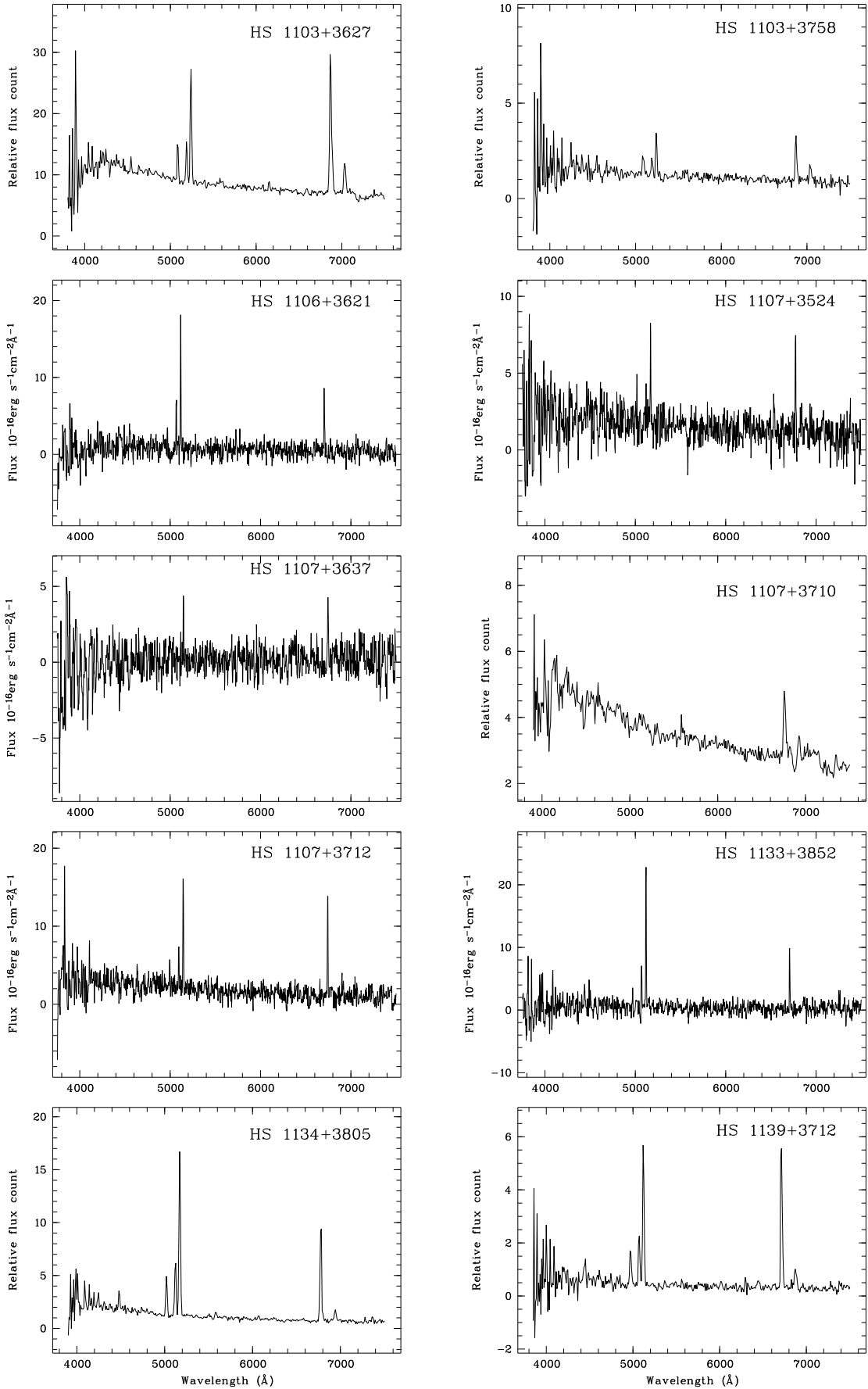
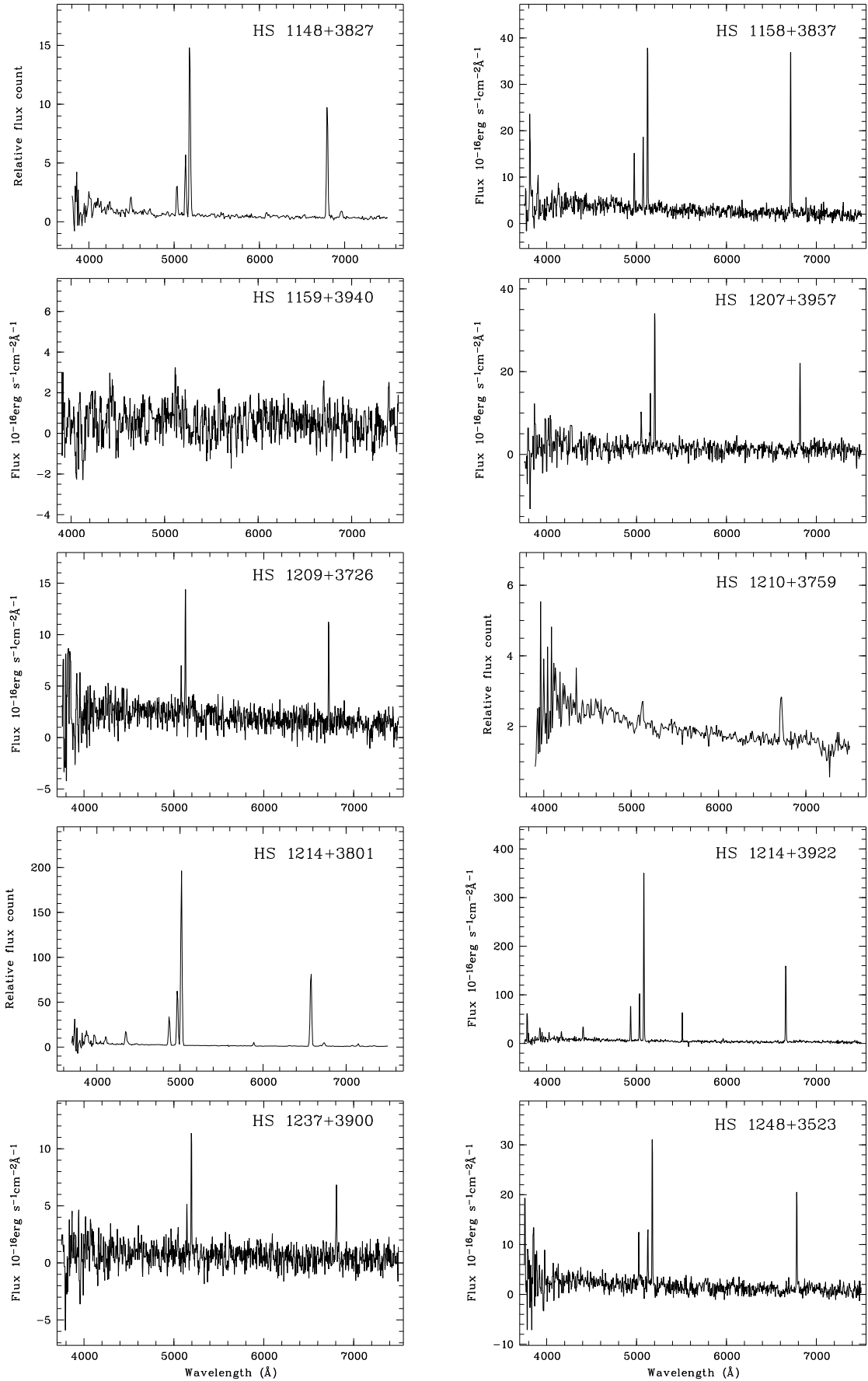
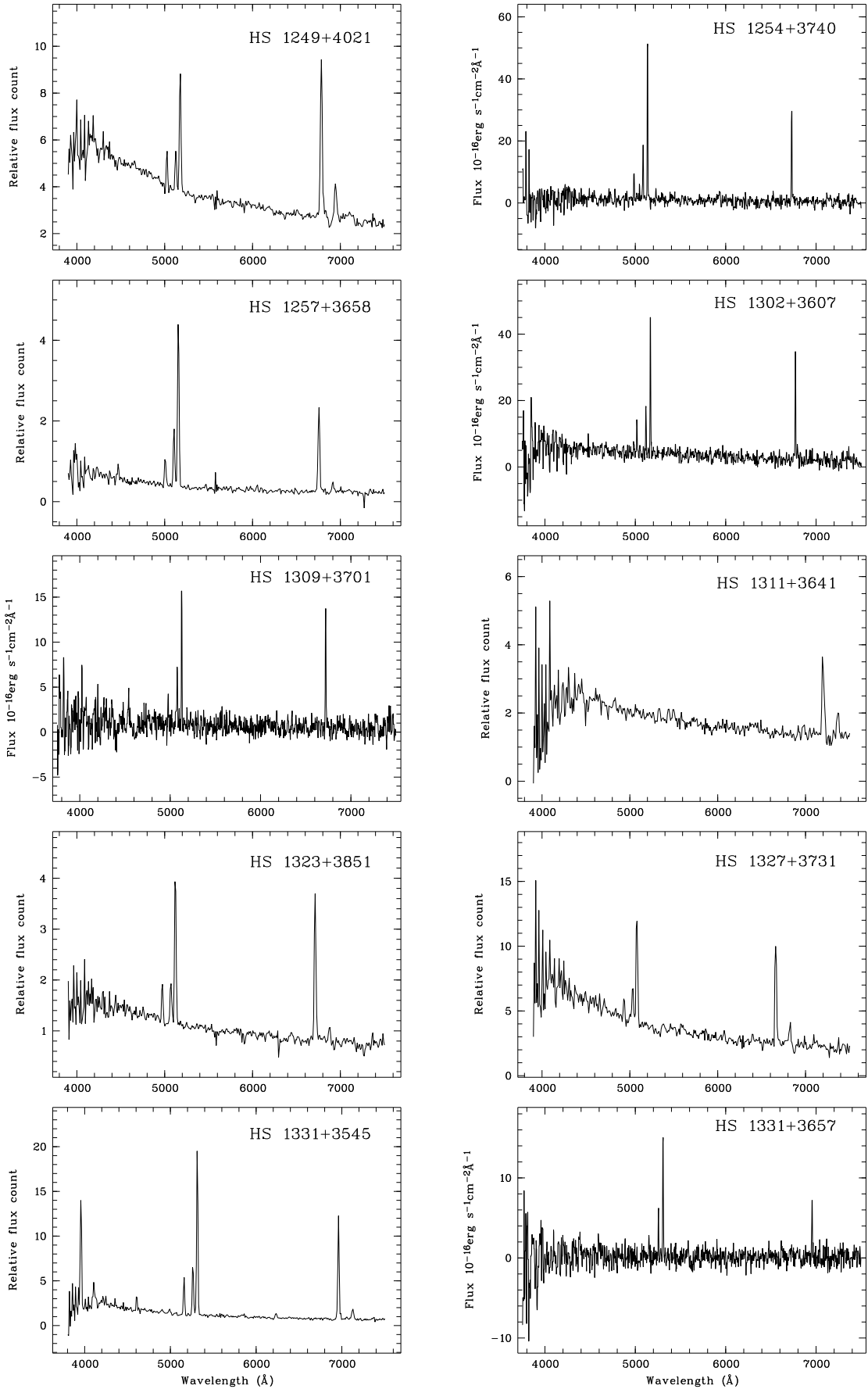
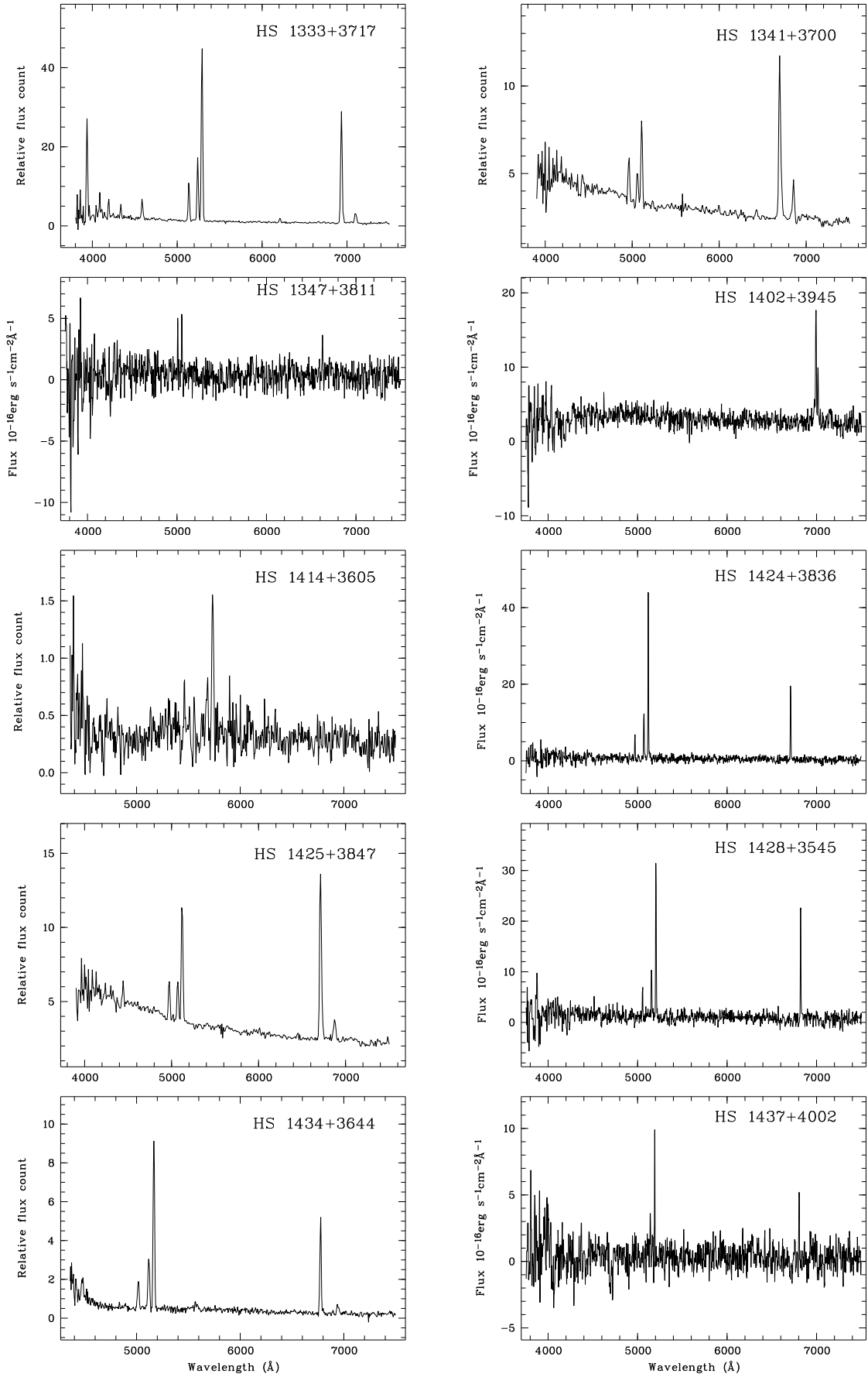
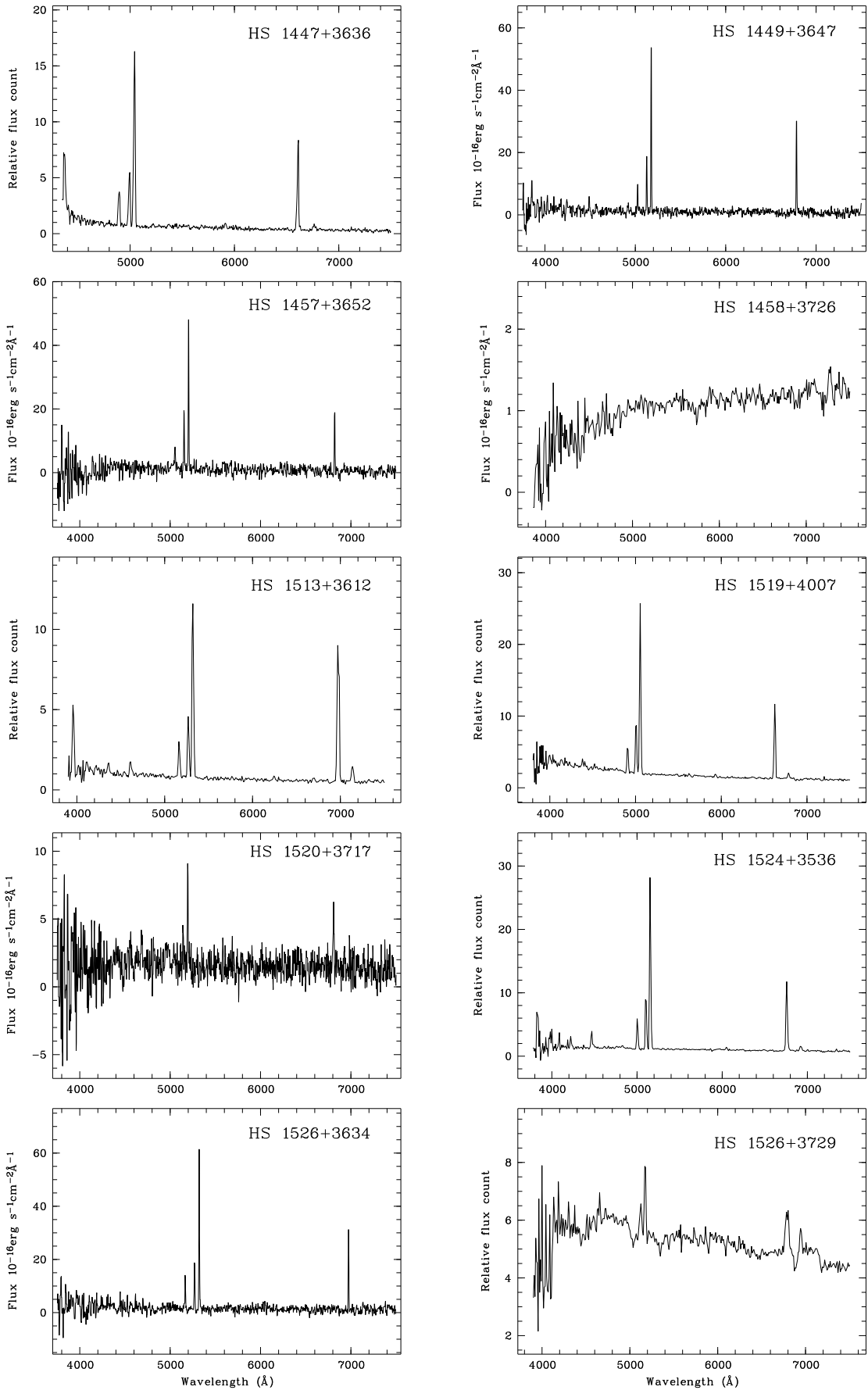
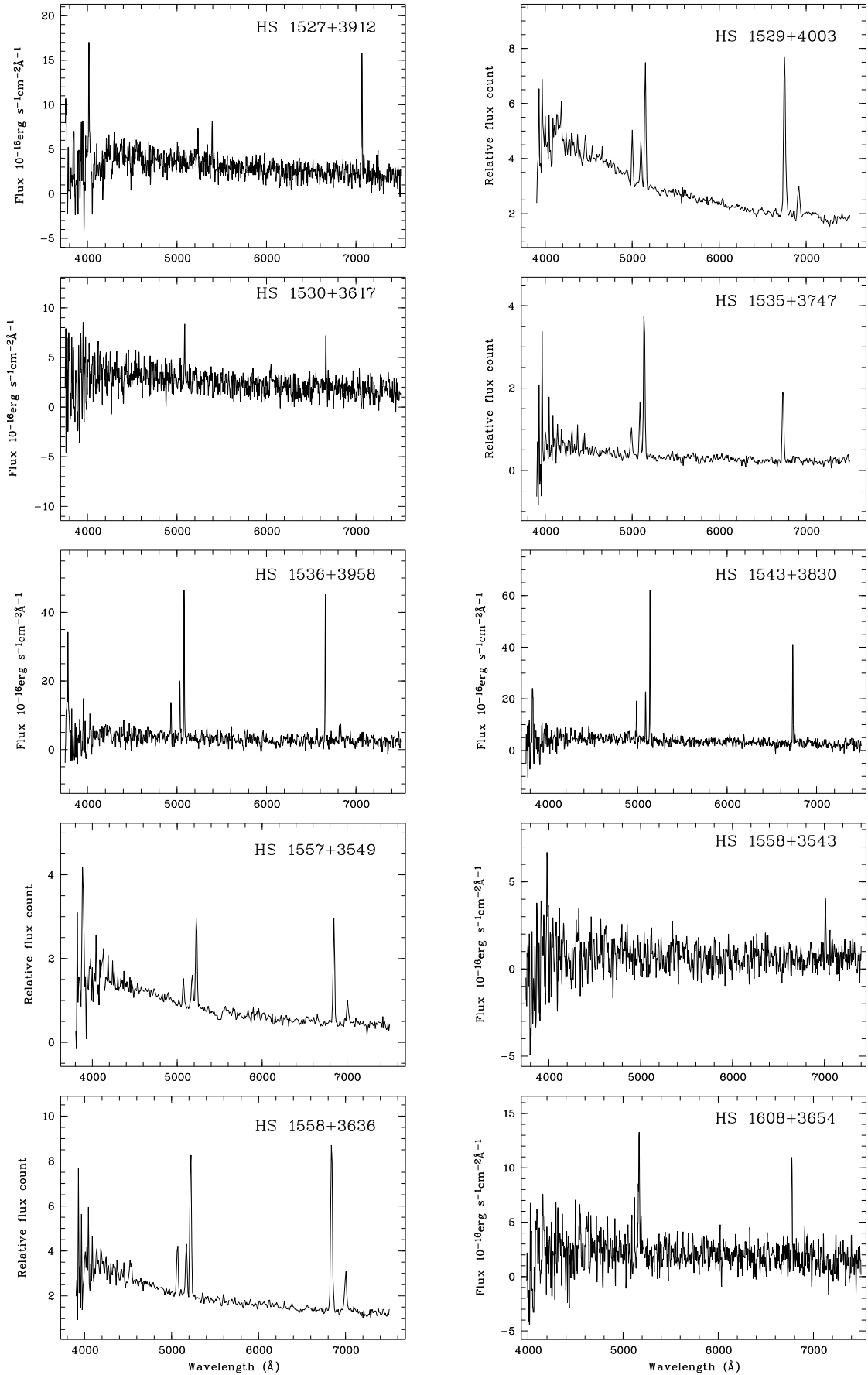
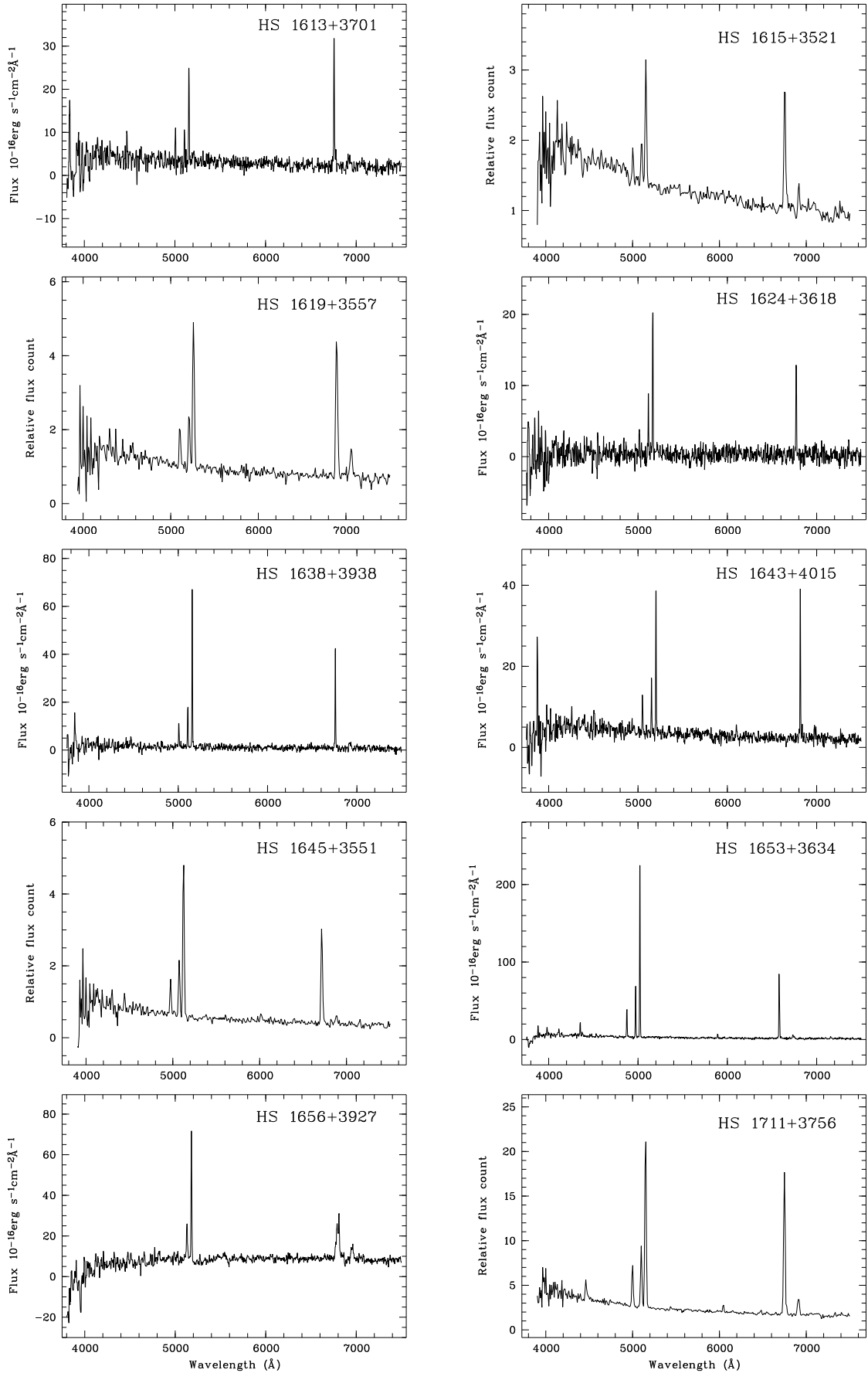
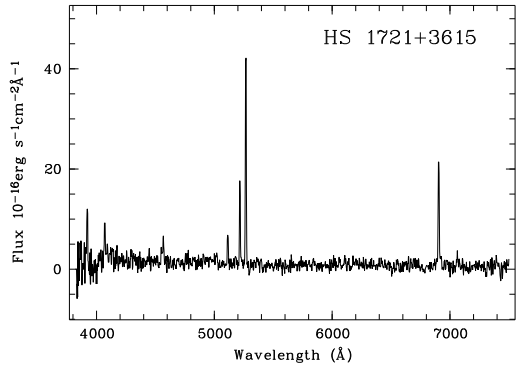
\end{document}